\documentclass[lettersize,journal]{IEEEtran}
\usepackage{amsmath,amsfonts}
\usepackage{algorithmic}
\usepackage{algorithm}
\usepackage{array}
\usepackage[caption=false,font=normalsize,labelfont=sf,textfont=sf]{subfig}
\usepackage{textcomp}
\usepackage{stfloats}
\usepackage{url}
\usepackage{verbatim}
\usepackage{graphicx}
\usepackage{cite}
\usepackage{multirow}
\usepackage{acronym}
\usepackage{booktabs}
\usepackage{pifont}
\usepackage{bm}
\usepackage{tabularx}
\usepackage{threeparttable}
\usepackage{xcolor}
\usepackage{caption}
\usepackage{makecell}
\usepackage[switch]{lineno}

\usepackage{xcolor}    
\usepackage{colortbl}

\usepackage[commandnameprefix=always,final]{changes}
 \usepackage{tikz}
 \newcommand{\circledstar}{%
 	\mathbin{%
 		\tikz[baseline=(C.base)] \node[draw,circle,inner sep=.7pt] (C) {$\scriptstyle*$};%
 	}%
 }

\begin{document}

\title{Task-Guided Prompting for Unified Remote Sensing Image Restoration}            


\author{Wenli Huang*, Yang Wu*, Xiaomeng Xin, Zhihong Liu, Jinjun Wang, and Ye Deng,~\IEEEmembership{Member,~IEEE}  
\thanks{Manuscript was received on September 9 2025; revised December 16 2025. Date of current version October 24 2025. This work was supported by the Ningbo key Research and Development Plan Project under Grant 2023Z230, the “Innovation Yongjiang 2035” Major Application Demonstration Programme under Grant 2024Z003, the Fundamental Research Funds for the Sichuan Science Foundation project under Grant 2024ZDZX0002 and Grant 2024NSFTD0054, and the Public Welfare Research Program of Ningbo City under Grant 2024S063.
	 (Corresponding author: Ye Deng.)}
\thanks{
	*These authors contributed equally to this work.
	
	Wenli Huang is with the School of Electronic and Information Engineering, Ningbo University of Technology, Ningbo, Zhejiang 315211, China (e-mail: huangwenli@nbut.edu.cn).
	
	Yang Wu, Xiaomeng Xin, and Jinjun Wang are with the Institute of Artificial Intelligence and Robotics, Xi’an Jiaotong University, China (e-mail: wuyang\_cc@stu.xjtu.edu.cn; jinjun@mail.xjtu.edu.cn).
	
	Zhihong Liu is with the University of Exeter, Exeter EX4 4PY, UK.
	
	Ye Deng is with the Engineering Research Center of Intelligent Finance, Ministry of Education, and the School of Computing and Artificial Intelligence, Southwestern University of Finance and Economics, China (e-mail: dengye@swufe.edu.cn).

}}

\markboth{~Vol.~XX, No.~XX, Dec.~2025}%
{Huang \MakeLowercase{\textit{et al.}}: Task-Guided Prompting for Unified Remote Sensing Image Restoration}

\IEEEpubid{0000--0000/00\$00.00~\copyright~2025 IEEE}

\maketitle

\begin{abstract}

Remote sensing image restoration (RSIR) is essential for recovering high-fidelity imagery from degraded observations, enabling accurate downstream analysis. However, most existing methods focus on single degradation types within homogeneous data, restricting their practicality in real-world scenarios where multiple degradations often across diverse spectral bands or sensor modalities, creating a significant operational bottleneck. 
To address this fundamental gap, we propose TGPNet, a unified framework capable of handling denoising, cloud removal, shadow removal, deblurring, and SAR despeckling within a single, unified architecture.
The core of our framework is a novel Task-Guided Prompting (TGP) strategy. TGP leverages learnable, task-specific embeddings to generate degradation-aware cues, which then hierarchically modulate features throughout the decoder. This task-adaptive mechanism allows the network to precisely tailor its restoration process for distinct degradation patterns while maintaining a single set of shared weights.
To validate our framework, we {construct a unified RSIR benchmark covering RGB, multispectral, SAR, and thermal infrared modalities for five} aforementioned restoration tasks.
Experimental results demonstrate that TGPNet achieves state-of-the-art performance on both unified multi-task scenarios and unseen composite degradations, surpassing even specialized models in individual domains such as cloud removal. 
By successfully unifying heterogeneous degradation removal within a single adaptive framework, this work presents a significant advancement for multi-task RSIR, offering a practical and scalable solution for operational pipelines.
The code and benchmark will be released at \url{https://github.com/huangwenwenlili/TGPNet}.

\end{abstract}


\section{Introduction}

\begin{figure}[!t]
	\centering
	\includegraphics[width=3.3in]{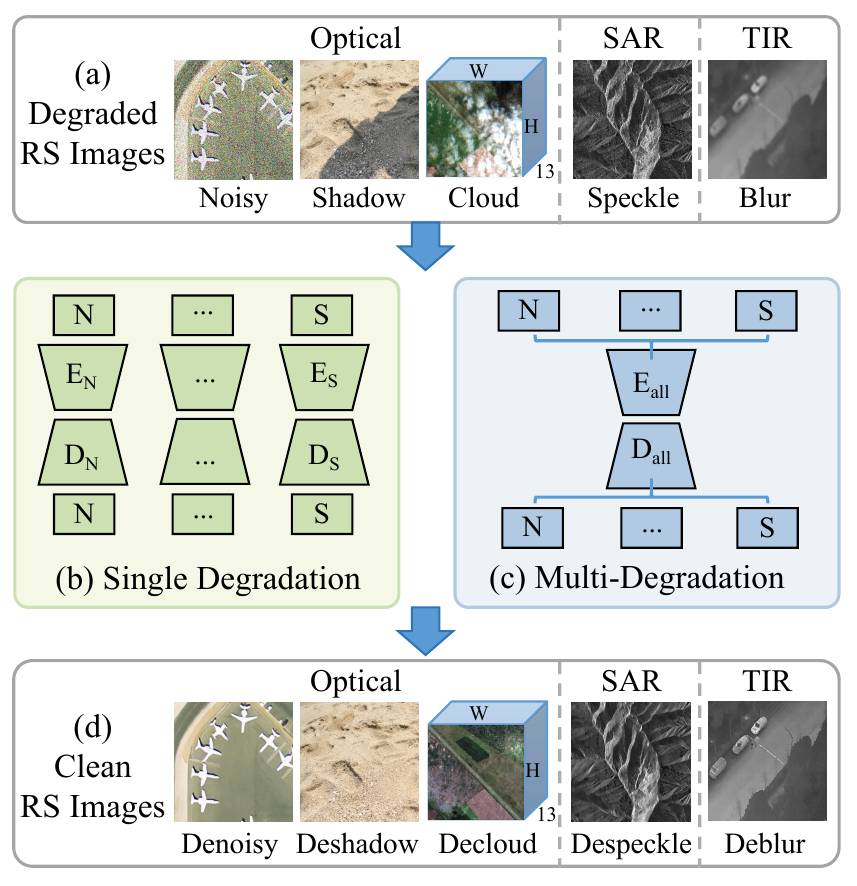}
	\caption{		
		Conceptual comparison of different RSIR paradigms. (a) Examples of real-world degradations across diverse modalities, including optical (RGB/Multispectral), Synthetic Aperture Radar (SAR), and Thermal Infrared (TIR). (b) Conventional single-task restoration relies on specialized networks (e.g., $E_N, D_N$ for denoising) for each degradation type. (c) Our unified approach employs a single, prompt-guided network ($E_{\text{all}}, D_{\text{all}}$) to handle diverse tasks, including denoising, deshadowing, declouding, despeckling, and deblurring. (d) The corresponding high-quality restored images.
	}
	\label{fig:urs_intro}
\end{figure}

\IEEEPARstart{R}{emote} sensing (RS) images are indispensable to modern Earth observation, underpinning critical applications in environmental monitoring \cite{casagli2023landslide}, resource management \cite{sheffield2018satellite}, and sustainable development \cite{holloway2018statistical}. 
However, the quality of raw imagery is frequently compromised by a variety of degradations. These range from atmospheric interference, such as clouds and shadows, to sensor-specific artifacts like noise and Synthetic Aperture Radar (SAR) speckle, which severely impair the accuracy of derived geospatial information (see Fig. \ref{fig:urs_intro}(a)). 
\IEEEpubidadjcol
Consequently, remote sensing image restoration (RSIR) has emerged as a crucial post-processing step aimed at reconstructing high-fidelity imagery from corrupted observations. Effective restoration not only improves visual quality but also significantly enhances the performance of downstream tasks, including object detection~\cite{zakria2022multiscale} and land-cover segmentation~\cite{kotaridis2021remote}.

Recent advances in deep learning have led to a proliferation of specialized models for single-task RSIR, which achieve impressive performance on isolated degradations such as denoising \cite{wang2024rsid}, declouding \cite{wu2024cr,liu2024cascaded}, shadow removal \cite{guo2023shadowformer}, deblurring \cite{xiao2024homoformer}, and SAR despeckling \cite{ko2021sar,fang2024contrastive}. 
While powerful, these task-specific approaches operate as “expert" models, each meticulously engineered for a single purpose (Fig. \ref{fig:urs_intro}(b)). Their practical applicability is severely constrained by the complex and dynamic nature of real-world remote sensing data, where multiple degradation types often vary spatiotemporally.
This inherent variability makes it infeasible to select and deploy an appropriate specialized model for every scenario, creating a significant bottleneck for operational RSIR pipelines.

The limitations of single-task models have motivated the development of unified restoration frameworks, primarily within the natural image domain~\cite{li2022all,potlapalli2306promptir,cuiadair}. However, these “all-in-one" models are not directly transferable to the remote sensing context due to the pronounced domain gap. 
Existing methods often treat restoration tasks in isolation or strictly within the optical domain. However, real-world remote sensing faces dual challenges: diverse degradation types (e.g., noise, clouds, shadows, blur, speckle) and heterogeneous imaging domains (e.g., optical RGB, SAR microwave, TIR thermal). These domains differ fundamentally in imaging physics (active vs. passive sensing) and degradation characteristics (additive vs. multiplicative noise), making unified restoration a non-trivial problem. Therefore, our work focuses on \textbf{Unified Cross-Domain Multi-Degradation Restoration}. We aim to design a single framework adaptable to both varied degradation objectives and distinct imaging domains. We validate TGPNet through a progressive experimental hierarchy: starting with two and three optical-domain tasks (intra-domain baseline), extending to include SAR despeckling (dual-domain generalization), and finally incorporating TIR deblurring (tri-domain comprehensive unification). This design systematically verifies the framework’s scalability across fundamental sensing modalities.

To address this critical need, we propose TGPNet, a unified framework designed to handle heterogeneous remote sensing image restoration (RSIR) tasks within a single model. Crucially, rather than embedding rigid, hand-crafted physical equations—which inherently conflict when unifying incompatible modalities like SAR speckle and optical degradations—TGPNet relies on \textbf{data-driven implicit priors}.
At the heart of TGPNet is our novel Task-Guided Prompting (TGP) strategy, which enables the network to perform diverse restoration tasks through a unified architecture (Fig. \ref{fig:urs_intro}(c)). TGP consists of two key components: a Learnable Task-Specific Embedding (LTSE) module that encodes the restoration task into a compact vector, and a Hierarchical Feature Modulation (HFM) module.
Functioning as a dynamic physical context switch, the HFM leverages this embedding to generate multi-scale affine parameters that modulate features across the decoder stages.
This mechanism allows the network to adaptively reconfigure feature statistics to match specific signal distributions, aligning its behavior with distinct degradation characteristics without altering its core weights. This effectively enables a single model to function as multiple experts.

Furthermore, to facilitate rigorous evaluation and advance future research, we construct a comprehensive multi-degradation benchmark for RSIR. This benchmark encompasses RGB, multispectral (MS), SAR, and thermal infrared (TIR) data across five key restoration tasks: denoising, cloud removal, shadow removal, deblurring, and SAR despeckling. Extensive evaluations demonstrate that TGPNet achieves state-of-the-art performance; it not only outperforms existing unified frameworks but also exhibits robust generalization to unseen composite degradations (e.g., simultaneous denoising and cloud removal) and surpasses specialized single-task models in their respective domains.

This work advances remote sensing image restoration (RSIR) through three key innovations:
\begin{itemize}
	\item We propose TGPNet, the first framework to unify the restoration of heterogeneous degradations across fundamentally distinct sensor modalities (optical, SAR, and TIR) within a single, prompt-guided network.
	\item We devise the Task-Guided Prompting (TGP) strategy, where a learnable task embedding guides a hierarchical modulation module to achieve dynamic, task-adaptive feature processing in the decoder.
	\item We construct and will publicly release a multi-modal benchmark for multi-degradation RSIR, on which TGPNet achieves state-of-the-art performance and demonstrates superior cross-task generalization.
\end{itemize}

\section{Related Works}

This section reviews the literature relevant to Remote Sensing Image Restoration (RSIR). We first survey methods developed specifically for RS data, noting a prevailing trend toward specialized models that lack a unified framework. We then analyze unified restoration models from the natural image domain to highlight their limitations for RSIR scenarios.

\subsection{Remote Sensing Image Restoration}
Remote sensing imagery is inherently susceptible to various degradations that hinder downstream analysis tasks like classification and change detection. These are commonly categorized into four types \cite{rasti2021image}: atmospheric interference (e.g., clouds), illumination effects (e.g., shadows), imaging artifacts (e.g., SAR speckle), and sensor noise. This susceptibility necessitates the development of robust image restoration methods to restore image quality.

Early RSIR methods were based on classical image processing techniques, such as spatial filtering, transforms (e.g., Fourier, wavelet \cite{shen2022coupling}), and regression analysis \cite{xu2018deep}. These approaches were often combined with regularization priors, including total variation and sparse representation, to constrain the solution. They struggled to model the complex, non-linear physics of degradation processes. Consequently, their performance was often suboptimal, particularly in scenarios with severe or mixed degradations.

The advent of deep learning has revolutionized RSIR, spawning a wave of state-of-the-art models—yet the field remains constrained by two critical limitations that hinder practical, unified deployment in multi-modal RS workflows.

{First, hyper-specialization to single tasks or modalities defines most current RSIR models—a paradigm that prevents generalization across diverse RS scenarios.}
Most deep learning-based RSIR models are narrowly optimized for one degradation type or sensor modality, as evidenced by their limited adaptability to cross-modal or multi-degradation tasks. This specialization is distinct across optical and SAR imagery.

For optical RS imagery, models are tailored to specific degradations, even when leveraging advanced design principles like multi-scale feature modulation. For large-scale, non-local phenomena (e.g., clouds, haze), transformer-based architectures such as CR-former \cite{wu2024cr} and cascaded model \cite{liu2024cascaded} excel at capturing long-range spatial dependencies but cannot address shadows, noise, or non-optical data. For localized issues, shadow removal relies on style guidance or specialized attention (e.g., ShadowFormer \cite{guo2023shadowformer}, HomoFormer \cite{xiao2024homoformer}) but cannot handle speckle or deblurring. Denoising models like HCANet \cite{hu2024hybrid} and RSID-CR \cite{wang2024rsid} are tightly coupled to additive noise distributions, making them incompatible with SAR’s speckle.

{Notably, even advanced multi-scale architectures—designed to capture hierarchical degradation patterns—fall into this specialization trap. Examples include MB-TaylorFormer V2 \cite{jin2025mb} (single-task restoration), DDMSNet \cite{zhang2021deep} (snow removal), ESTINet \cite{zhang2022enhanced} (video deraining), DBLRNet \cite{zhang2018adversarial} (video deblurring), and LLdiffusion \cite{wang2025lldiffusion} (low-light enhancement). None of these models can handle the diverse degradations inherent to static, multi-modal RS data.}

For SAR imagery, the unique statistical properties of speckle (a result of coherent imaging) demand entirely separate solutions. Models like DeSpeckNet \cite{mullissa2020despecknet} (CNN-based suppression) and CL-SAR-Despeckling \cite{fang2024contrastive} (contrastive learning for texture preservation) are specialized for speckle removal but cannot process optical degradations (e.g., clouds, shadows) or share weights across modalities. This forces users to deploy disjoint model suites for multi-modal RS workflows.

{Second, recent attempts to address multi-degradation RSIR have resulted in pseudo-unified systems—frameworks that appear versatile but rely on task-specific components or separate training, rather than a single shared backbone.} Examples include HIRDiff \cite{pang2024hir} requires independent training for each restoration task (e.g., denoising vs. deshadowing), undermining scalability. HDI-PRNet \cite{feng2024progressive} employs distinct task-specific modules (e.g., cloud-specific encoders, shadow-specific decoders), increasing computational overhead and limiting adaptability. PromptHSI \cite{lee2024prompthsi}, an RS-focused prompt-based model restricted to hyperspectral imagery—it cannot handle cross-modal tasks (e.g., optical + SAR) or thermal deblurring.
Despite their progress, these methods are not true all-in-one solutions: they either require reconfiguration for new tasks or fail to span optical/SAR modalities.

{These two limitations create a defining gap in the RSIR landscape: no existing method enables a single model to handle cross-optical/SAR modalities and multiple degradations (cloud, shadow, speckle, noise, deblur) without task-specific retraining or components. This gap motivates our Task-Guided Prompting (TGP) module, which integrates task-specific prompts into multi-scale feature modulation to empower a single shared backbone with dynamic adaptation. In doing so, TGPNet positions itself as the first truly unified cross-optical/SAR multi-task RSIR framework.}

\subsection{Unified Image Restoration}

To overcome the limitations of task-specific models, unified frameworks for multi-degradation restoration have emerged—with foundational research primarily from the natural image domain. These frameworks are grouped into five paradigms: architectural unification, transformer-based modeling, diffusion-based refinement, language-guided adaptation, and prompt-based control \cite{jiang2024survey}.

Early unified frameworks prioritized modular architectures to balance task specificity and computational efficiency. IPT \cite{chen2021pre} introduced a multi-head/multi-tail design, enabling degradation-adaptive feature processing via a shared backbone; AirNet \cite{li2022all} enhanced this by integrating contrastive learning to improve cross-degradation feature discrimination; LoRA-IR \cite{ai2024lora} proposed a flexible framework with low-rank experts guided by a degradation-aware router, achieving state-of-the-art performance across 14 tasks and 29 benchmarks; and MOCE-IR \cite{zamfir2025complexity} developed a mixture-of-experts (MoE) architecture with variable computational complexity and receptive fields for universal degradation restoration.

Transformer architectures have been adapted to model complex composite degradations. Such as, OneRestore \cite{guo2024onerestore} uses cross-attention to fuse degraded scene descriptors for joint multi-degradation restoration, while AllRestorer \cite{mao2024allrestorer} integrates transformer blocks to process multi-modal inputs (e.g., image-text pairs) and improve degraded scene understanding. Diffusion models leverage iterative denoising for multi-degradation restoration (though with higher computational overhead). Such as, WeatherDiff \cite{ozdenizci2023restoring} uses conditional diffusion for weather-related degradations; MPerceiver \cite{ai2024multimodal} and AutoDIR \cite{jiang2024autodir} enhance diffusion via multimodal guidance (e.g., textual/visual prompts) and structural-corrected latent diffusion; and UniRestore \cite{chen2025unirestore} and UniCoRN \cite{mandal2025unicorn} integrate perceptual and task-oriented features into diffusion priors for controllable multi-degradation handling.

Multimodal large language models (MLLMs) and prompt-based methods offer additional unified solutions. Such as, MLLM-based works (e.g., Q-Agent \cite{zhou2025q}, VLU-Net \cite{zeng2025vision}) enable degradation-aware guidance and flexible restoration control via vision-language alignment; prompt-based methods (e.g., InstructIR \cite{conde2024instructir}, SPIRE \cite{qi2024spire}) provide lightweight control, with later works like MPMF-Net \cite{wen2025multi} and AdaIR \cite{cuiadair} further advancing this paradigm.

Despite demonstrating promising performance on natural image datasets (e.g., ImageNet), these unified frameworks exhibit critical limitations when applied to remote sensing (RS) scenarios: insufficient modeling of degradation processes tailored to sensor-specific noise characteristics, inadequate spectral guidance mechanisms for multi-band RS data restoration, and notable domain discrepancies between natural RGB images and RS data (e.g., multi-spectral, synthetic aperture radar (SAR) modalities). This underscores the need for domain-adaptive solutions capable of addressing heterogeneous RS degradations across diverse spectral bands—an essential requirement for practical operational deployment in RS applications.

\section{Methodology}
This section details the proposed Task-Guided Prompting Network (TGPNet), a unified framework for multi-degradation restoration of remote sensing imagery. We first present the overall architecture of TGPNet, which integrates Task-Guided Prompting (TGP) modules within an encoder-decoder structure. Next, we elaborate on the core TGP module, which learns degradation-specific characteristics and modulates decoder features to enable fine reconstruction across diverse degradations. Finally, we outline the optimization objective.

\begin{figure*}[tbp]
	\centering
	\includegraphics[width=7.0in]{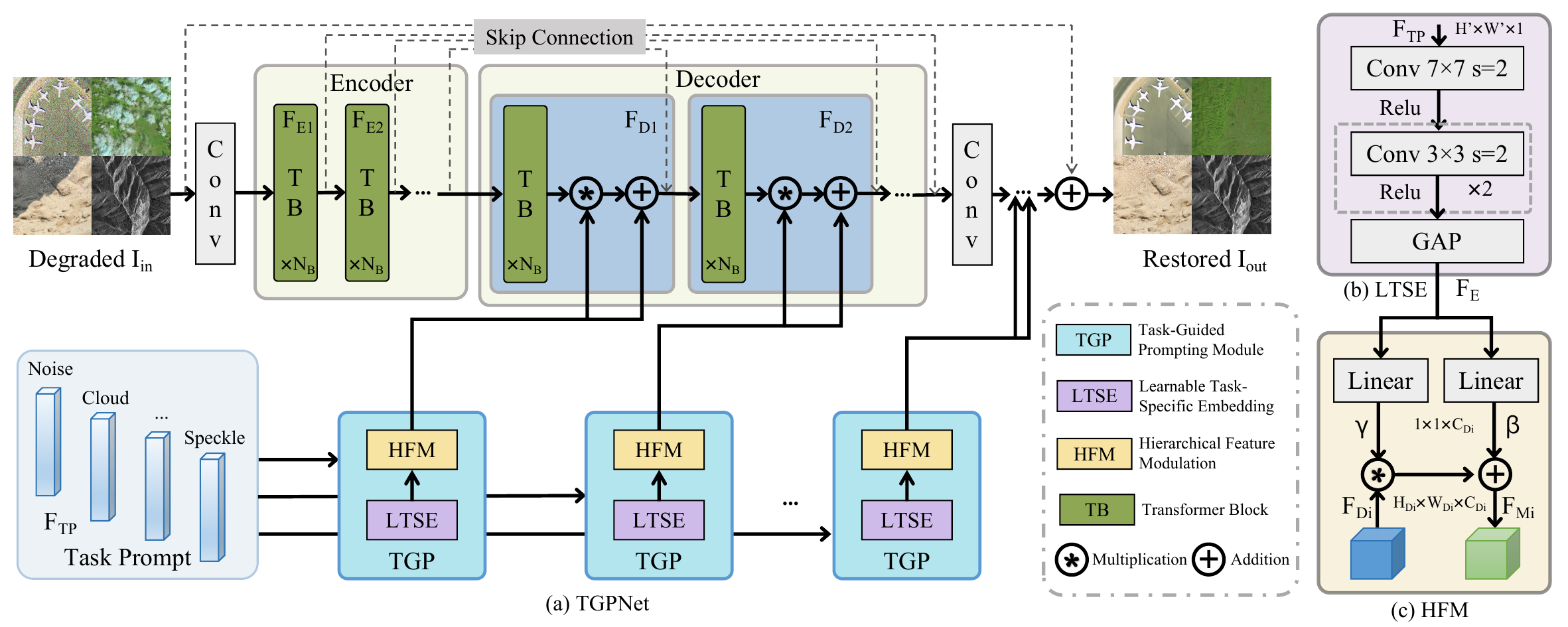}
	\caption{Architecture of the proposed Task-Guided Prompting Network (TGPNet) for unified remote sensing image restoration. (a) The overall architecture consists of an encoder-decoder backbone augmented by our Task-Guided Prompting (TGP) module, which is composed of Learnable Task-Specific Embedding (LTSE) and Hierarchical Feature Modulation (HFM). (b) The LTSE module processes a task-specific prompt ($F_{TP}$) to produce a degradation-aware embedding ($F_{E}$). (c) The HFM module uses this embedding to generate stage-wise affine parameters (scaling factors $\gamma$ and shifting offsets $\beta$), which modulate the hierarchical features ($F_{Di}$) at each stage of the decoder.
	}
	\label{fig:network_architecture}
\end{figure*}

\subsection{TGPNet Architecture}

TGPNet is designed to learn a unified mapping function $\mathcal{M}$ that reconstructs a clean image $I_{\text{out}}$ from a degraded input $I_{\text{in}}$, conditioned on a task prompt $F_{\text{TP}}$. It is formally defined as:
\begin{equation}
I_{\text{out}} = \mathcal{M}(I_{\text{in}}, F_{\text{TP}}; \Theta),
\end{equation}
where $F_{\text{TP}}$ denotes the task prompt encoding degradation characteristics in $I_{\text{in}}$, and $\Theta$ represents the learnable parameters of the network.

As illustrated in Fig. \ref{fig:network_architecture}(a), the TGPNet backbone is a U-shaped encoder-decoder structure adapted from Restormer \cite{zamir2022restormer}. The core building blocks are transformers that use multi-head transposed self-attention across channel dimensions for long-range dependency modeling, coupled with gated depthwise convolutions in the feed-forward network for local feature refinement. The encoder hierarchically downsamples the input through four stages, progressively reducing spatial resolution while increasing channel depth. Symmetrically, the decoder upsamples the features through four stages, with skip connections concatenating features from corresponding encoder-decoder levels to preserve spatial details. Crucially, our Task-Guided Prompting (TGP) modules are integrated into each decoder stage to enable degradation-aware reconstruction.

\subsubsection{Encoder}
Given a degraded remote sensing input image $I_{in} \in \mathbb{R}^{H \times W \times C_{in}}$, where $H$, $W$, and $C_{in}$ denote the height, width and channel, respectively. The encoder first projects it into a higher-dimensional feature space using a $3 \times 3$ convolution, yielding shallow features
extracts shallow {high-dimensional} features $F_{S} \in \mathbb{R}^{H \times W \times C}$, where $C$ denotes the basic channel count. 
Following this initial step, $F_{S}$ is passed through four hierarchical stages to extract multi-scale representations.  Each stage $i \in {1,2,3,4}$ refines the features using a series of Transformer Blocks (TBs) and then performs spatial downsampling via pixel unshuffling. This sequential process generates four sets of feature maps, $F_{Ei}$, where the spatial resolution is halved and the channel dimension is doubled at each stage, resulting in features of size $\mathbb{R}^{H/2^{(i-1)} \times W/2^{(i-1)} \times 2^{(i-1)}C}$.


\subsubsection{Decoder}
Symmetrically to the encoder, the decoder reconstructs the high-resolution image from the deep features, $F_{E4}$, across four sequential stages. Each stage begins by upsampling the features using a pixel shuffle operation, which doubles the spatial resolution and halves the channel count. The upsampled features are then concatenated with the corresponding multi-scale features, $F_{Ei}$, from the encoder via skip connections. Following channel alignment with a $1 \times 1$ convolution, the fused features are refined by a series of Transformer Blocks. After the final stage, the features are passed through a refinement module and a terminal $3 \times 3$ convolution to generate residual image, $I_r \in \mathbb{R}^{H \times W \times C_{in}}$, which is added to the original input to produce the final restored image: $I_{out} = I_{in} + I_r$.

The core innovation of our network is the Task-Guided Prompting (TGP) module, which enables degradation-aware reconstruction by adjusting decoder features based on task-specific embeddings. To achieve this hierarchical control, the TGP module is strategically deployed at six key points in the reconstruction pathway: four within individual decoder stages (specifically, post-Transformer blocks and pre-upsampling), one after the refinement module, and one after the terminal convolution layer. This multi-scale, targeted modulation allows a single, unified model to effectively handle heterogeneous degradations. The operational details of the TGP module and its components are elaborated in the following section.





\subsection{Task-Guided Prompting (TGP)}
To enable a unified network to handle heterogeneous degradations across diverse spectral bands or sensor modalities (Fig. \ref{fig:urs_intro}), we introduce the Task-Guided Prompting (TGP) module. The TGP module adapts the restoration process by operating through two synergistic components: a Learnable Task-Specific Embedding (LTSE) module and a Hierarchical Feature Modulation (HFM) module. First, the LTSE processes a task prompt to generate a compact, degradation-aware embedding. This embedding then guides the HFM module, which generates and applies multi-scale affine transformations to hierarchically modulate the decoder features. This two-stage process allows the network to precisely tailor its restoration strategy to the specific degradation at hand. The design of each component is detailed below.





\subsubsection{Learnable Task-Specific Embedding (LTSE)}

The LTSE module is designed to distill a task prompt, $F_{TP}  \in \mathbb{R}^{H' \times W' \times 1}$, into a compact, degradation-aware feature vector. 
This vector serves as a latent physical descriptor, implicitly encoding the distinct characteristics of sensor-specific degradations.
As illustrated in Fig. \ref{fig:network_architecture}(b), the LTSE, which is jointly optimized with the main restoration network, processes the randomly initialized task prompt through a hierarchical convolutional architecture. This architecture consists of three sequential convolutional layers with ReLU activations, which progressively downsample the spatial dimensions. This hierarchical design allows the module to abstract degradation characteristics from coarse to fine scales. Finally, a global average pooling (GAP) layer compresses the spatial information into a final degradation-specific embedding, $F_{E} \in \mathbb{R}^{1 \times 1 \times C_E}$. This entire operation is formally expressed as:
\begin{equation}
F_{E} = \text{GAP}\left(\sigma\left(W_3 \left(\sigma\left(W_2 \left(\sigma\left(W_1 \left(F_{TP}\right)\right)\right)\right)\right)\right)\right),
\end{equation}
where $W_1$ is a $7 \times 7$ convolutional kernel and $W_2, W_3$ are $3 \times 3$ convolutional kernels, all with a stride of 2 for downsampling. $\sigma$ denotes the ReLU activation.


\subsubsection{Hierarchical Feature Modulation (HFM)}

The HFM module uses the degradation embedding $F_{E}$ to adaptively refine features $F_{Di} \in \mathbb{R}^{H_{Di} \times W_{Di} \times C_{Di}}$ at each decoder stage, refinement module, and terminal convolution. As depicted in Fig. \ref{fig:network_architecture}(c), it first generates specific affine parameters \cite{perez2018film}—a scale factor $\gamma$ and a shift factor $\beta$—from the embedding using two distinct linear layers:
\begin{equation}
\gamma =  \text{Linear}_1(F_E),\quad \beta =  \text{Linear}_2(F_E).
\end{equation}
To enable channel-wise modulation, the linear layers are designed such that their output dimension matches the feature channel count, $C_{Di}$. This count is determined by the stage's depth, where $C_{Di} = 2^{(i-1)}C$ and $C$ is the network's base channel dimension.

The generated affine parameters, $\gamma$ (scale) and $\beta$ (shift), are then applied to the features $F_{Di}$, to produce the modulated features $F_{Mi}$. This is achieved via a affine transformation:
\begin{equation}
F_{Mi} = \gamma \circledstar F_{Di} + \beta,
\end{equation}
where $\circledstar$ denotes channel-wise multiplication. This operation adaptively recalibrates the decoder feature maps $F_{Di}$, aligning the network's internal representations with the specific signal characteristics of the target degradation type.

The TGPNet architecture is strategically designed to integrate task-specific guidance at multiple scales. This is achieved by deploying six TGP modules at key points in the network.
At each location, the two-step TGP mechanism—comprising the LTSE and HFM—generates and applies scale-aware affine parameters to task-adaptively modulate the feature maps. This hierarchical and dynamic modulation bridges the gap between the general-purpose backbone and specific task requirements, enabling a single, unified model to achieve precise reconstruction of heterogeneous degradations.



\subsection{Objective Function}

The $\mathcal{L}_1$ loss is adopted as the primary optimization objective for training TGPNet, leveraging its demonstrated efficacy in preserving edge details and minimizing reconstruction artifacts in image restoration tasks \cite{zhao2016loss}. This loss function quantifies pixel-wise discrepancies between the restored output $I_{\text{out}}$ and the ground truth reference $I_{\text{gt}}$, defined as the mean absolute error (MAE) across all spatial positions and spectral channels:
\begin{equation}
\mathcal{L}_1 = \frac{1}{HWC} \sum_{h=1}^{H} \sum_{w=1}^{W} \sum_{c=1}^{C} |I_{\text{out}}^{(h,w,c)} - I_{\text{gt}}^{(h,w,c)}|,
\end{equation}
where $H$, $W$, and $C$ denote the image height, width, and channel count, respectively. The normalization factor $\frac{1}{H W C}$ computes the expectation per pixel-channel element, enabling scale-invariant optimization. This formulation ensures consistent gradient behavior across varying input resolutions while preserving fine details.






\section{Experiments}

This section presents the experimental validation of TGPNet. We begin by describing the benchmark datasets, experimental setting, and evaluation metrics. We then conduct a comprehensive performance analysis, assessing TGPNet in two distinct scenarios: first as a unified framework handling four heterogeneous degradation types, and second as a specialized model for cloud removal. To conclude, we perform a series of ablation studies to quantify the contribution of our core architectural components and validate key design choices.


\begin{table}[t]
	\centering
	\caption{Overview of the datasets comprising the benchmark for unified remote sensing image restoration (URSIR).}
	\resizebox{.99\columnwidth}{!}  {
		\begin{tabular}{m{7.5em}<{\centering}m{8em}<{\centering}m{2.2em}<{\centering}m{1.8em}<{\centering}m{6em}<{\centering}}
			\toprule
			Task  & Dataset & Train & Test & Resolution \\
			\midrule
			Denoising  & UCMLUD \cite{yang2010bag} & 1680  & 420   & $256 \times 256$  \\
			\midrule
			\multirow{3}[0]{*}{Declouding} & RICE1 \cite{lin2019remote} & 400 & 100 & $512 \times 512$   \\
			& RICE2 \cite{lin2019remote} & 558   & 148  & $512 \times 512$  \\
			& {SEN12MS-CR} \cite{meraner2020cloud} & 114,319   & 7,899  & $256 \times 256$  \\
			\midrule
			\multirow{2}[0]{*}{Deshadowing} & SRD \cite{qu2017deshadownet}   & 2680  & 408 & $640 \times 480$   \\
			& UAV-TSS  & 2,200   & 400  & $1920 \times 1080$  \\
			\midrule
			SAR Despeckling & NRD \cite{guan2023robust}   & 250   & 20 &  $512 \times 512$   \\
			\midrule
			Deblurring & {HIT-UAV} \cite{suo2023hit}  & 2,029   & 579 &  $640 \times 512$   \\
			
			\bottomrule
		\end{tabular}%
	}
	\label{tab:benchmark}%
\end{table}%

\subsection{Benchmark}

To facilitate the evaluation of unified remote sensing image restoration (URSIR) models, we establish a comprehensive benchmark designed to test performance across a wide range of tasks, data modalities, and imaging conditions. This benchmark integrates five distinct restoration challenges: declouding, deshadowing, denoising, deblurring, and SAR despeckling. The included datasets span multiple sensor types—RGB, multispectral, SAR, and thermal infrared—to ensure that evaluated models are robust and versatile enough for real-world applications.

The benchmark is composed of the following standard datasets:

\begin{itemize}
	\item Denoising: The UC Merced Land Use Dataset (UCMLUD) \cite{yang2010bag} is utilized for additive Gaussian noise removal, featuring high-resolution optical imagery from 21 diverse land-use categories.
	\item Declouding: Three datasets are included to address declouding across various cloud types and sensor modalities. The RICE1 and RICE2 \cite{lin2019remote} datasets provide RGB imagery for removing thin and thick clouds, respectively. {The SEN12MS-CR \cite{meraner2020cloud} dataset offers a multispectral challenge, featuring 13-band Sentinel-2 imagery with varied cloud cover.}	
	\item Deshadowing: Shadow correction is evaluated using two distinct sources. First, the Shadow Removal Dataset (SRD) \cite{qu2017deshadownet} provides real-world optical images featuring complex shadow geometries across urban and natural landscapes. Second, we constructed a specialized UAV Traffic Surveillance Scenes (UAV-TSS) dataset (2,600 images), which utilizes linear illumination modeling to simulate realistic aerial occlusion and penumbra effects.
	\item SAR Despeckling: The Near Real Dataset (NRD) \cite{guan2023robust} is used for speckle reduction in SAR imagery. It consists of speckled Sentinel-1 data and corresponding clean reference images generated from time-series analysis.
	\item {Deblurring: The HIT-UAV dataset \cite{suo2023hit} is used for Gauss deblurring, presenting a challenging task with thermal infrared imagery captured by UAVs in low-contrast environments.}

\end{itemize}
Detailed specifications for each component dataset are provided in Table \ref{tab:benchmark}.

\begin{table}[t]
	\centering
	\caption{Detail of our TGPNet architecture.}
	\resizebox{.9\columnwidth}{!}  {
		\begin{tabular}{m{4.6em}<{\centering}m{10.7em}<{\centering}m{14em}<{\centering}}
			\toprule
			Stage & Operator & Param \& Output  \\
			\midrule
			Input & -     & H$\times$W$\times$ $C_{in}$ \\
			\midrule
			Stem & Conv$(3\times C_{in}\times C)$ & H$\times$W$\times C$ \\
			\midrule
			$E_1$ & [TB] $\times 1$ & $N_h=2$, H$\times$W$\times C$ \\
			\midrule
			Downsample & Conv, PixelUnshuffle(2) & H/2$\times$W/2$\times$ $2C$  \\
			\midrule
			$E_2$ & [TB] $\times 2$ &$N_h=4$,  H/2$\times$W/2$\times2C$  \\
			\midrule
			Downsample  & Conv, PixelUnshuffle(2) & H/4$\times$W/4$\times4C$  \\
			\midrule
			$E_3$ & [TB] $\times 2$  &$N_h=8$,  H/4$\times$W/4$\times4C$ \\
			\midrule
			Downsample  & Conv, PixelUnshuffle(2) & H/8$\times$W/8$\times8C$  \\
			\midrule
			$E_4$ & [TB] $\times 4$  &$N_h=8$,  H/8$\times$W/8$\times8C$ \\
			\midrule			
			$D_4$ & [TB] $\times 4$ &  $N_h=8$,  H/8$\times$W/8$\times8C$ \\
			\midrule						
			TGP & LTSE, HFM &  $ \gamma \in \mathbb{R}^{1 \times 8C}$,  $\beta \in \mathbb{R}^{1 \times 8C}$, H/8$\times$W/8$\times 8C$ \\
			\midrule
			Upsample  & Conv, PixelShuffle(2), Cat, Reduce\_Ch & H/4$\times$W/4$\times$ $4C$  \\
			\midrule
			$D_3$ & [TB] $\times 2$ & $N_h=8$,  H/4$\times$W/4$\times4C$ \\
			\midrule			
			TGP & LTSE, HFM &  $ \gamma \in \mathbb{R}^{1 \times 4C}$,  $\beta \in \mathbb{R}^{1 \times 4C}$, H/4$\times$W/4$\times 4C$ \\
			\midrule
			Upsample  & Conv, PixelShuffle(2), Cat, Reduce\_Ch& H/2$\times$W/2$\times$ $2C$  \\
			\midrule
			$D_2$ & [TB] $\times 2$ & $N_h=4$,  H/2$\times$W/2$\times$ $2C$ \\
			\midrule
			TGP & LTSE, HFM &  $ \gamma \in \mathbb{R}^{1 \times 2C}$,  $\beta \in \mathbb{R}^{1 \times 2C}$, H/2$\times$W/2$\times 2C$ \\
			\midrule
			Upsample  & Conv, PixelShuffle(2), Cat & H$\times$W$\times$ $2C$  \\
			\midrule			
			$D_1$ & [TB] $\times 1$ &$N_h=2$, H$\times$W$\times$$2C$ \\
			\midrule
			TGP & LTSE, HFM &  $ \gamma \in \mathbb{R}^{1 \times 2C}$,  $\beta \in \mathbb{R}^{1 \times 2C}$, H$\times$W$\times 2C$ \\
			\midrule
			Refinement & [TB] $\times 2$ &$N_h=2$, H$\times$W$\times$$2C$ \\
			\midrule
			TGP & LTSE, HFM &  $ \gamma \in \mathbb{R}^{1 \times 2C}$,  $\beta \in \mathbb{R}^{1 \times 2C}$, H$\times$W$\times 2C$ \\
			\midrule
			Residual generation & Conv$(3\times C\times C_{in})$ & H$\times$W$\times$ $C_{in}$ \\
			\midrule
			TGP & LTSE, HFM &  $ \gamma \in \mathbb{R}^{1 \times C_{in}}$,  $\beta \in \mathbb{R}^{1 \times C_{in}}$, H$\times$W$\times C_{in}$ \\
			\midrule
			Output & Add Input and residual feature & H$\times$W$\times$$C_{in}$ \\
			\bottomrule
		\end{tabular}
	}	
	
	\begin{tablenotes}
		\footnotesize
		\item[*] $C_{in}$ denotes the number of input channels, $C$ represents the base number of channels. A convolutional filter $Conv(k \times C_{1} \times C_{2})$ is defined with a kernel size of $k$, input and output channels counts of $C_{1}$ and $C_{2}$. $N_h$ indicates the number of attention heads. TB refers to the Transformer block. For modulating features of the corresponding encoder layer, $\gamma$ serves as the scaling factor and $\beta$ as the shifting factor.
	\end{tablenotes}
	
	\label{tab:net-details}%
\end{table}%

\subsection{Experimental Setting}

To enable unified restoration across diverse tasks, we employ a joint training strategy. Specifically, we consolidate datasets from all five degradation tasks into a single training pool. In each iteration, a training batch is constructed by randomly drawing samples from this unified pool. Based on the degradation type of each sample, the corresponding task-specific prompt is retrieved to modulate the input features, allowing the model to optimize for all tasks simultaneously.

We trained TGPNet for 450 epochs using a batch size of 14 on two NVIDIA RTX 4090 GPUs. The AdamW optimizer \cite{loshchilov2017decoupled} was employed with $\beta_1=0.9$, $\beta_2=0.999$. The learning rate was managed by a cosine annealing schedule with restarts, beginning with an initial rate of $2 \times 10^{-4}$ held constant for the first 90 epochs to ensure stable initial learning. This was followed by two consecutive 180-epoch cycles where the learning rate was annealed down to $1 \times 10^{-6}$. We also utilized an Exponential Moving Average (EMA) with a decay of $\alpha=0.999$ to further stabilize training. During training, the model was fed with $128 \times 128$  patches, randomly cropped from the source images and augmented with random flips and rotations. The base channel dimension was set to $C=$48, with further architectural details provided in Table \ref{tab:net-details}.


\subsection{Evaluation Metrics}

For quantitative evaluation, we use a set of standard metrics to assess restoration quality. Across all tasks in our unified RSIR benchmark, we employ the peak signal-to-noise ratio (PSNR) to measure reconstruction fidelity and the structural similarity index measure (SSIM) to evaluate perceptual quality \cite{cuiadair, jiang2024survey}. For the specialized declouding task, we follow established practice by also including the mean absolute error (MAE) for pixel-wise accuracy and the spectral angle mapper (SAM) to assess spectral preservation \cite{wu2024cr, huang2024ACAcrnet}. 


\begin{table*}[htbp]
  \centering
  \caption{Comparisons for unified remote sensing image restoration (RSIR) across two and three degradation types (PSNR/SSIM; higher = better; best in bold). }
  \resizebox{.99\textwidth}{!}  {
   \begin{tabular}{m{6.8em}<{\centering}m{4.3em}<{\centering}m{4.3em}<{\centering}m{4.3em}<{\centering}m{4.3em}<{\centering} c m{4.3em}<{\centering}m{4.3em}<{\centering}m{5em}<{\centering}m{4.5em}<{\centering}c}
    \toprule
    \multirow{3}[6]{*}{Method} & \multicolumn{5}{c}{2 Types}           & \multicolumn{5}{c}{3 Types} \\ \cmidrule(r){2-6}\cmidrule(r){7-11}     
         & \multicolumn{2}{c}{Declouding} & \multicolumn{2}{c}{Deshadowing} & \multirow{2}[4]{*}{Average} & \multicolumn{2}{c}{Declouding} & \multicolumn{1}{l}{Deshadowing} & Denoising & \multirow{2}[4]{*}{Average} \\
\cmidrule(r){2-3}\cmidrule(r){4-5} \cmidrule(r){7-8}  \cmidrule(r){9-9} \cmidrule(r){10-10}              & on RICE1 & on RICE2 & on SRD & on UAV-TSS &       & on RICE1 & on RICE2 & on SRD & Avg. on UCMLUD &  \\
    \cmidrule(r){1-1}\cmidrule(r){2-6}\cmidrule(r){7-11}     
    PromptIR \cite{potlapalli2306promptir}  & 30.32/0.9445 & 33.36/0.8960 & 24.77/0.8559 & 40.39/0.9911 & 32.21/0.9219 & 31.24/0.9502 & 32.57/0.8973 & 24.90/0.8570 & 31.20/0.8630 & 29.98/0.8919 \\
    AdaIR \cite{cuiadair} & 28.98/0.9417 & 33.65/0.8966 & 25.02/0.8563 & 39.97/0.9909 & 31.91/0.9214 & 24.28/0.9047 & 29.37/0.8817 & 23.62/0.8507 & 31.00/0.8596 & 27.07/0.8742 \\
    MOCE-IR \cite{zamfir2025complexity} & 30.98/0.9441 & 31.69/0.8844 & 24.17/0.8509 & 39.85/0.9869 & 31.67/0.9166 & 32.75/0.9509 & 32.77/0.8941 & 24.63/0.8562 & 31.04/0,8601 & 30.30/0.8903 \\
    PromptHSI \cite{lee2024prompthsi} & 32.22/0.9513 & 33.56/0.8888 & 24.89/0.8563 & 38.67/0.9831 & 32.34/0.9199 & 30.65/0.9507 & 33.81/0.8915 & 24.52/0.8512 & 23.57/0.6406 & 28.14/0.8335 \\
    Art \cite{wu2024harmony}  & 32.26/0.9308 & 31.79/0.8520 & 24.46/0.8399 & 30.10/0.9574 & 29.65/0.8950 & 28.51/0.8954 & 28.75/0.8244 & 22.86/0.8060 & 26.73/0.7465 & 26.71/0.8181 \\
    CR-former \cite{wu2024cr} & 35.01/0.9569 & 35.49/0.9075 & 27.13/0.8679 & 48.82/0.9958 & 36.61/0.9320 & 35.07/0.9579 & 35.66/0.9095 & 27.24/0.8696 & 31.47/0.8704 & 32.36/0.9019 \\
    \cmidrule(r){1-1}\cmidrule(r){2-6}\cmidrule(r){7-11}     
    Ours  & \textbf{35.70/0.9603} & \textbf{35.80/0.9115} & \textbf{28.00/0.8727} & \textbf{52.72/0.9969} & \textbf{38.06/0.9354} & \textbf{35.92/0.9606} & \textbf{36.05/0.9130} & \textbf{27.96/0.8732} & \textbf{31.65/0.8737} & \textbf{32.90/0.9051} \\
    \bottomrule
    \end{tabular}%
}
  \label{tab:performance-task2-task3}%
\end{table*}%

\begin{table*}[htbp]
	\centering
	\caption{Comparisons for unified remote sensing image restoration (RSIR) across four degradation types (PSNR/SSIM; higher = better; best in bold). }
	\resizebox{.99\textwidth}{!}  {
		\begin{tabular}{m{7em}<{\centering}cccccccc}
			\toprule
			\multirow{2}[2]{*}{Method} & \multicolumn{2}{c}{Declouding} & \multicolumn{1}{l}{Deshadowing} &  \multicolumn{3}{c}{Denoising on UCMLUD} & \multicolumn{1}{l}{SAR Despeckling} & \multirow{2}[2]{*}{Average} \\
            \cmidrule(r){2-3}  \cmidrule(r){4-4} \cmidrule(r){5-7} \cmidrule(r){8-8} & on RICE1 & on RICE2 & on SRD &  $\sigma=15$ & $\sigma=25$ & $\sigma=50$ &   On NRD &    \\
			\midrule
			PromptIR \cite{potlapalli2306promptir} & 32.38/0.9539 & 34.67/0.9067 & 26.33/0.8653 &  34.07/0.9224 & 31.78/0.8827 & 28.82/0.8102 & 24.91/0.8311 & 30.42/0.8818  \\
			AdaIR \cite{cuiadair} & 33.10/0.9566 & 35.12/0.9094 & 26.99/0.8679 & 34.14/0.9238 & 31.85/0.8844 & 28.89/0.8124 & 25.05/0.8262 & 30.73/0.8830   \\
			MOCE-IR \cite{zamfir2025complexity} & 31.85/0.9561 & 33.88/0.9105 & 26.14/0.8685 & 34.13/0.9240 & 31.83/0.8844 & 28.86/0.8113 & \textbf{25.15}/0.8310 & 30.26/0.8838  \\
            PromptHSI \cite{lee2024prompthsi} & 32.15/0.9522 &34.22/0.9010 &25.96/0.8610  &24.94/0.8034 &24.59/0.7058 &21.57/0.4719 &24.79/0.8213 &26.89/0.7881 \\
            {Art \cite{wu2024harmony}} & 29.70/0.9183 & 30.73/0.8599 & 23.65/0.8188 & 29.05/0.8505 & 28.01/0.7998 & 25.93/0.7028 & 21.31/0.6537 &26.91/0.8005 \\            
			CR-former \cite{wu2024cr} & 35.64/0.9608 & 35.71/0.9114 & 27.47/0.8706 & 34.08/0.9231 & 31.79/0.8837 & 28.84/0.8116 & 24.86/0.8187 & 31.20/0.8826 \\
			\midrule
			Ours  & \textbf{35.73/0.9608} & \textbf{36.35/0.9145} & \textbf{28.40/0.8747} &  \textbf{34.27/0.9255} & \textbf{31.97/0.8868} & \textbf{29.05/0.8169} & 24.88/\textbf{0.8329} & \textbf{31.52/0.8874}   \\
			\bottomrule
		\end{tabular}%
	}
	\label{tab:performance-4task}%
\end{table*}%

\begin{figure*}[!t]
	\centering
	\includegraphics[width=7in]{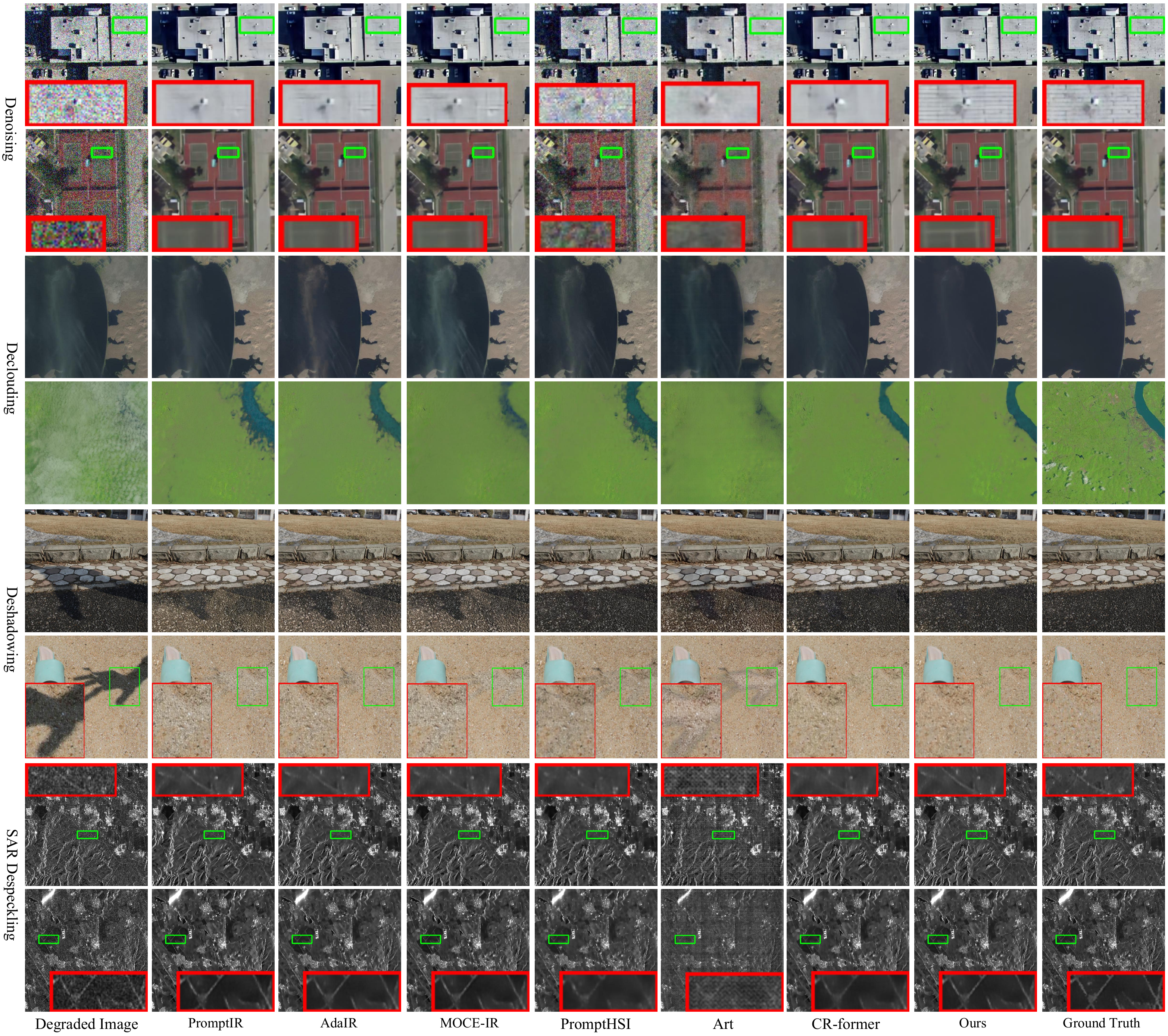}
	\caption{Visual comparison of restored images for four degradation types on our URSIR benchmark. Key local details (in green boxes) are highlighted in red boxes, and zooming in is recommended for detailed inspection.}
	\label{fig:performance-4task}
\end{figure*}

\begin{table*}[htbp]
	\centering
	\caption{{Comparisons for unified remote sensing image restoration (RSIR) across five degradation types (PSNR/SSIM; higher = better; best in bold).}}
	\resizebox{.99\textwidth}{!}  {
		\begin{tabular}{ccccccccc}
			\toprule
			\multirow{2}[2]{*}{Method} & \multicolumn{3}{c}{Declouding} & \multicolumn{1}{l}{Deshadowing} & Denoising & \multicolumn{1}{l}{SAR Despeckling} &  Deblurring & \multirow{2}[2]{*}{Average} \\
			\cmidrule(r){2-4}  \cmidrule(r){5-5} \cmidrule(r){6-6} \cmidrule(r){7-7} \cmidrule(r){8-8}         & On RICE1 & On RICE2 & On SEN12MS-CR & On SRD  & Avg. on UCMLUD & On NRD & Avg. on HIT-UAV &  \\
			\midrule
			PromptIR \cite{potlapalli2306promptir}& 32.30/0.9510 & 32.66/0.9025 & 28.73/0.8712 & 25.17/0.8592 & 31.06/0.8563 &24.97/0.8093 &  31.61/0.8701 & 29.50/0.8742 \\
			AdaIR  \cite{cuiadair}& 32.01/0.9517 & 33.72/0.9045 & 29.47/0.8789 & 25.53/0.8625 &  31.31/0.8656 & 25.10/0.8241 & 31.82/0.8722 & 29.85/0.8799 \\
			MOCE-IR \cite{zamfir2025complexity}& 29.40/0.9240 & 31.91/0.8956 & 29.07/0.8763 & 25.01/0.8596 & 30.85/0.8546 & 24.95/0.8121 &  31.26/0.8633 & 28.92/0.8694 \\
			Art \cite{wu2024harmony}  & 29.85/0.8414 & 30.71/0.8411 & 27.82/0.8599 & 23.92/0.8103 &  27.36/0.7732 & 20.96/0.6542 & 28.93/0.8244 & 27.08/0.8006 \\
			CR-former  \cite{wu2024cr}& 35.46/0.9588 & 35.85/0.9109 & 29.35/0.8926 & 27.53/0.8719 &  31.43/0.8701 & \textbf{25.31}/0.8273 & 32.26/0.8771 & 31.03/0.8870 \\
			\midrule
			Ours  & \textbf{35.59/0.9591} & \textbf{35.93/0.9130} & \textbf{29.61/0.8963} & \textbf{27.97/0.8739} &  \textbf{31.55/0.8723} & 24.95/\textbf{0.8294} & \textbf{32.35/0.8789} & \textbf{31.14/0.8890} \\
			\bottomrule
		\end{tabular}%
	}
	\label{tab:performance-5task}%
\end{table*}%

\begin{figure*}[!t]
	\centering
	\includegraphics[width=7in]{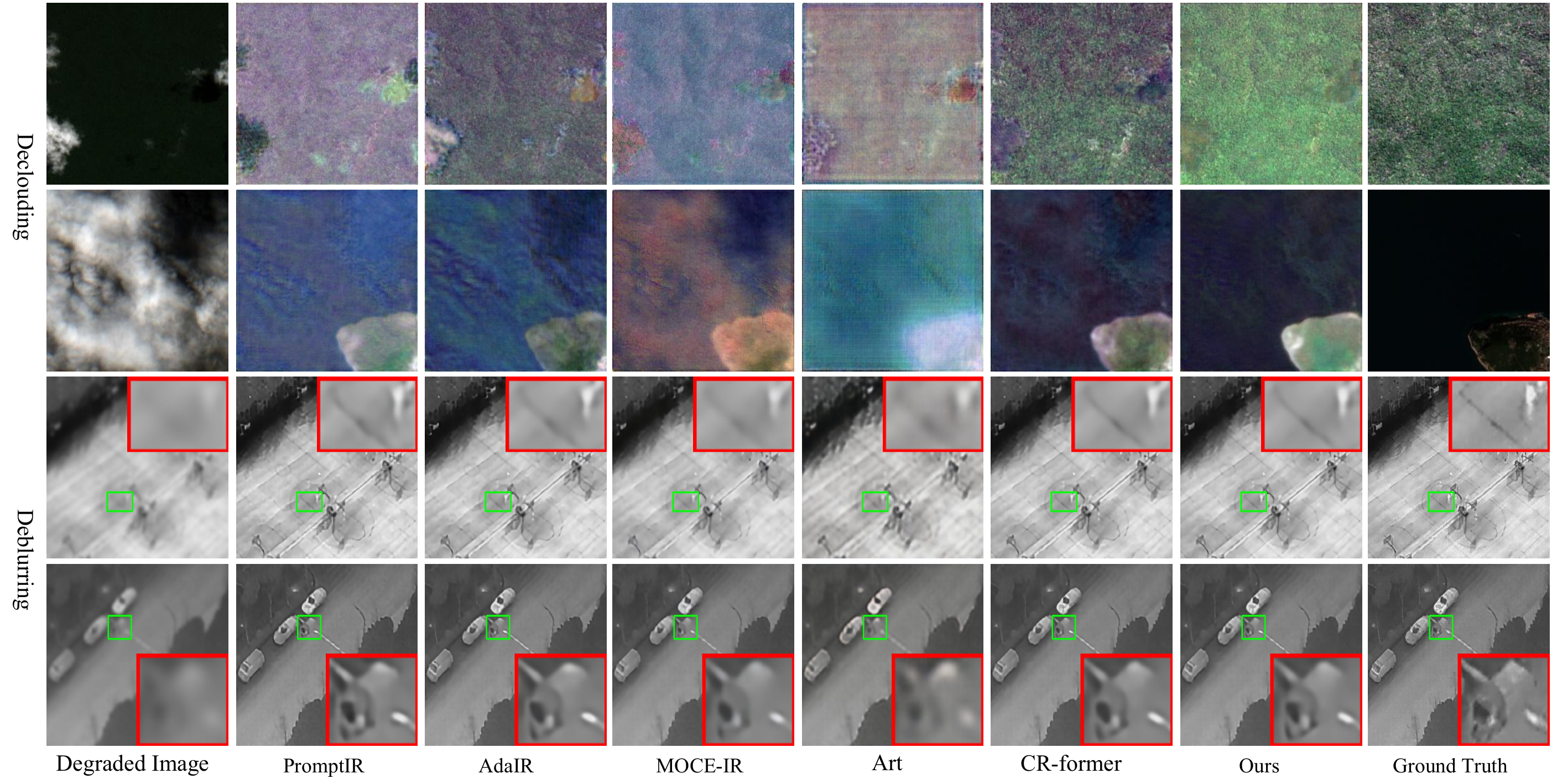}
	\caption{{Visual comparison of restored images for multispectral declouding on SEN12MS-CR and thermal deblurring on HIT-UAV. Key local details (in green boxes) are highlighted in red boxes, and zooming in is recommended for detailed inspection.}}
	\label{fig:performance-5task}
\end{figure*}

\begin{figure*}[!t]
	\centering
	\includegraphics[width=6.8in]{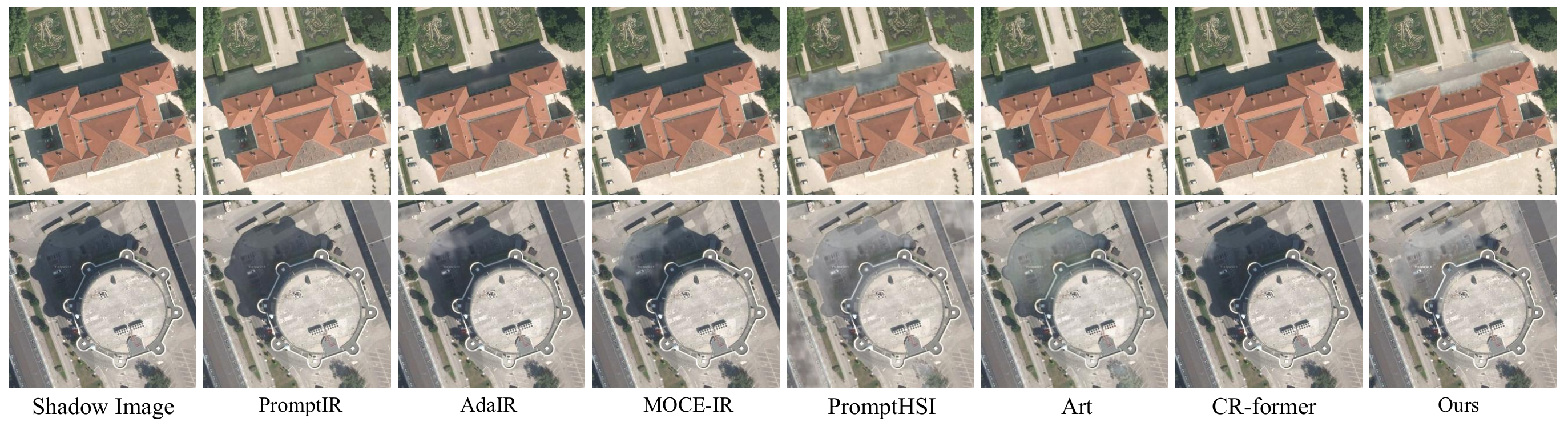}
	\caption{Visual evaluation of TGPNet on unseen real-world imagery from the WHU-Shadow dataset \cite{luo2020deeply}, demonstrating generalization to authentic remote sensing shadows.}
	\label{fig:deshadow-test}
\end{figure*}

\subsection{Performance on Diverse Restoration Tasks}\label{sec:comp-multi-degradations}

To rigorously evaluate TGPNet’s unified restoration capability and cross-domain scalability, we design a progressive experimental hierarchy with three settings, where each step systematically increases the complexity of domain heterogeneity and degradation diversity.

\subsubsection{Comparison Baselines}
To comprehensively evaluate TGPNet, we benchmark it against six state-of-the-art methods that span different domains and specializations. The selected baselines include four general-purpose natural image restoration models, one multi-degradation hyperspectral model, and one specialized cloud removal network. The models are:

\begin{itemize}
	\item PromptIR (NeurIPS 2023) \cite{potlapalli2306promptir}: A prompt-guided all-in-one restoration method that encodes degradation-specific information through learnable prompts, enabling multi-task processing for natural images.
	\item AdaIR (ICLR 2024) \cite{cuiadair}: An adaptive all-in-one natural IR method that mines spatial and frequency domain patterns to decouple heterogeneous degradations, facilitating restoration through domain-aware feature modulation.
	\item MOCE-IR (CVPR 2025) \cite{zamfir2025complexity}: A mixture-of-experts (MoE) architecture that selectively activates specialized experts for efficient multi-degradation handling, balancing computational complexity with restoration accuracy.
	\item PromptHSI (2025) \cite{lee2024prompthsi}: A prompt-guided multi-degradation restoration method for hyperspectral imagery using vision-language and frequency modulation.     
	\item {Art (ACMMM 2024) \cite{wu2024harmony}: An active re-weighting method designed to mitigate optimization conflicts in unified natural image restoration. }
	\item CR-former (TGRS 2024) \cite{wu2024cr}: A Transformer-based cloud removal method employing Focused-Taylor Attention to capture pixel-level long-range dependencies with linear computational complexity.
\end{itemize}

\subsubsection{Intra-Domain Optical Baseline (2 and 3 Tasks)}

We first establish a baseline using the Intra-Domain Optical setting, encompassing Denoising, Cloud Removal, and Shadow Removal. As detailed in Table \ref{tab:performance-task2-task3}, TGPNet demonstrates superior unified learning capability across both 2-task and 3-task configurations. It achieves the highest average PSNR and SSIM, significantly outperforming the runner-up CR-former \cite{wu2024cr}. This confirms that our prompt-guided mechanism effectively handles diverse degradations (noise, clouds, shadows) under consistent optical imaging physics.

\subsubsection{Dual-Domain Cross-Degradation (4 Tasks)}

Moving beyond the optical domain, we evaluate generalization to the Microwave Domain by incorporating SAR Despeckling. The quantitative results in Table \ref{tab:performance-4task} confirm TGPNet's robust cross-domain scalability despite the fundamental shift from optical noise to SAR speckle.

Specifically, TGPNet achieves the highest scores for cloud removal, outperforming the specialized CR-Former by 0.26 dB on RICE1 (35.73 dB) and 0.58 dB on RICE2 (36.35 dB). Notably, it surpasses the generalist PromptIR by a substantial 3.35 dB margin on RICE1. Furthermore, TGPNet maintains leading performance in shadow removal (SRD) and denoising (UCMLUD). Overall, it achieves the highest average metrics (31.52 dB / 0.8874), exceeding the next-best unified model, MOCE-IR, by 0.78 dB.

Fig. \ref{fig:performance-4task} visualizes this cross-domain success. In denoising, our approach achieves finer detail recovery, evidenced by enhanced reconstruction of roof textures and basketball court boundaries. For cloud removal, it effectively eliminates heterogeneous cloud cover while preserving river topology. In shadow removal, TGPNet accomplishes artifact-free restoration. Crucially, for the newly added SAR despeckling task, our network generates reconstructions with significantly enhanced edge acuity in terrain morphology (plains/mountains), proving its ability to adapt to coherent imaging characteristics.

\subsubsection{Tri-Domain Comprehensive Unification (5 Tasks)}
Finally, we evaluate the ultimate generalization capability by incorporating TIR Deblurring (Thermal Domain) to form the full 5-task benchmark. The results in Table \ref{tab:performance-5task} demonstrate that TGPNet successfully unifies restoration across Optical, Microwave, and Thermal domains without suffering from performance degradation on established tasks.

Notably, TGPNet excels on the newly added tasks, outperforming the strong competitor CR-former in both multispectral declouding on SEN12MS-CR (29.61 dB vs. 29.35 dB) and thermal deblurring on HIT-UAV (32.35 dB vs. 32.26 dB). Importantly, this extensibility comes with no trade-off in original task performance; TGPNet retains leading scores in RGB declouding, deshadowing, and denoising.

The superior quantitative performance is mirrored in the visual results (Fig. \ref{fig:performance-5task}). For the thermal deblurring task, the model recovers sharper edges and finer textures. In multispectral declouding, it effectively removes occlusion while preserving spectral integrity. Across all five tasks, TGPNet balances artifact suppression with detail preservation, validating its effectiveness as a robust, all-in-one restoration solution.

To further verify the robustness of our approach in unconstrained scenarios, we conducted qualitative evaluations on unseen, real-world imagery from the WHU-Shadow dataset \cite{luo2020deeply}. As visualised in Fig. \ref{fig:deshadow-test}, the results confirm the model's capability to remove authentic remote sensing shadows in scenarios not encountered during training, demonstrating effective generalization.

Collectively, these results confirm that our method not only achieves state-of-the-art quantitative and visual performance but also exhibits robust scalability across an expanding set of diverse restoration tasks and distinct imaging domains.

\begin{figure*}[!t]
	\centering
	\includegraphics[width=7in]{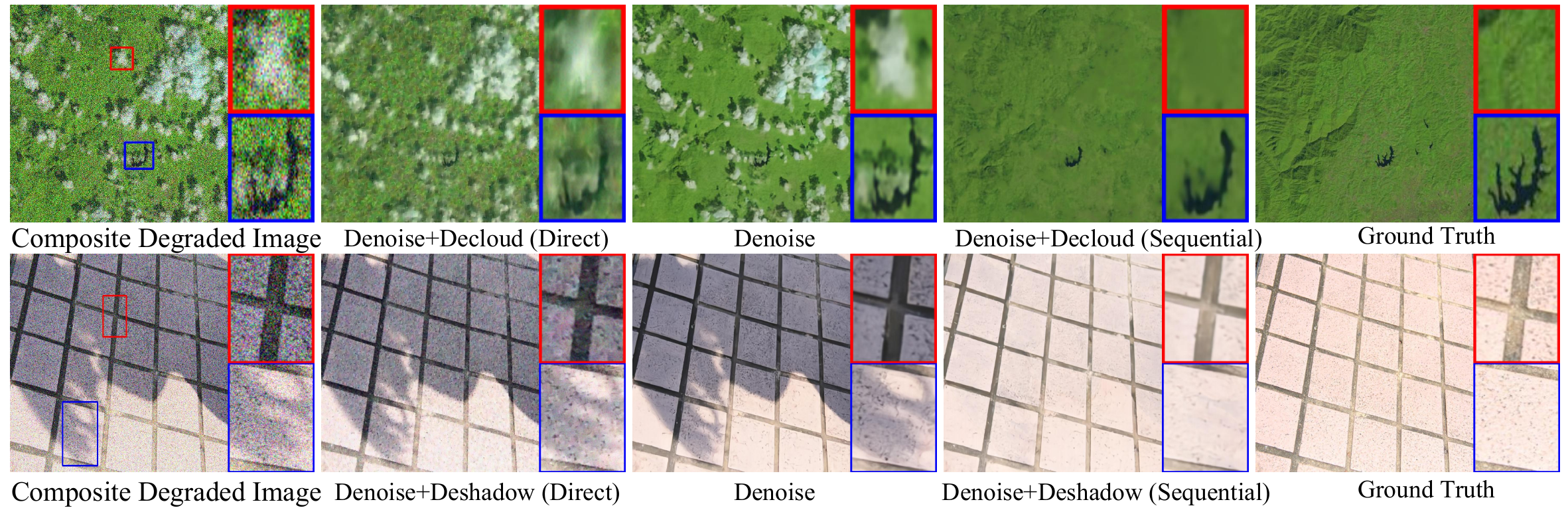}
	\caption{
		Visual comparison of restoration results for out-of-distribution composite degradations: direct vs. sequential processing. Direct approach (task embedding averaging) performs suboptimally. Sequential processing achieves effective restoration by decomposing composite degradations into ordered stepwise subtasks. Row 1: Denoising + Declouding; Row 2: Denoising + Deshadowing. Zoom in for details.
	}
	\label{fig:compose-step}
\end{figure*}

\begin{table*}[htbp]
  \centering
  \caption{Quantitative comparison on out-of-distribution two-composite-degradation tasks. Evaluated methods include RSIR baselines, our direct approach (task embedding averaging), and our sequential approach (ordered task-specific subtasks).}
  \resizebox{.99\textwidth}{!}  {
    \begin{tabular}{m{6.8em}<{\centering}cccccccccc}
    \toprule
    \multirow{2}[2]{*}{Method} & \multicolumn{3}{c}{Denoising+Declouding on RICE1} & \multicolumn{3}{c}{Denoising+Declouding on RICE2} & \multicolumn{3}{c}{Denoising+Deshadowing on SRD} & \multirow{2}[2]{*}{Average} \\
\cmidrule(r){2-4}  \cmidrule(r){5-7} \cmidrule(r){8-10}  
& $\sigma=15$ & $\sigma=25$ & $\sigma=50$ & $\sigma=15$ & $\sigma=25$ & $\sigma=50$ & $\sigma=15$ & $\sigma=25$ & $\sigma=50$ &  \\
    \midrule
    PromptIR \cite{potlapalli2306promptir} & 18.61/0.7062 & 18.47/0.6764 & 18.36/0.6312 & 22.32/0.7747 & 22.26/0.7709 & 22.22/0.7669 & 17.83/0.7622 & 17.70/0.7278 & 17.44/0.6583 & 19.47/0.7194 \\
    AdaIR \cite{cuiadair} & 18.54/0.7070 & 18.47/0.6754 & 18.34/0.6308 & 22.27/0.7730 & 22.27/0.7709 & 22.21/0.7676 & 17.85/0.7608 & 17.71/0.7282 & 17.46/0.6602 & 19.46/0.7193 \\
    MOCE-IR \cite{zamfir2025complexity} & 18.56/0.7129 & 18.47/0.6794 & 18.31/0.6307 & 22.31/0.7722 & 22.23/0.7693 & 22.16/0.7669 & 17.96/0.7622 & 17.72/0.7286 & 17.45/0.6597 & 19.46/0.7202 \\
    PromptHSI \cite{lee2024prompthsi} & 22.93/0.6268 & 20.02/0.4469 & 17.18/0.2411 & 30.06/0.8202 & 26.18/0.6869 & 20.61/0.3238 & 23.09/0.7046 & 20.66/0.5603 & 17.15/0.3320 & 21.99/0.5270 \\
    
    Art \cite{wu2024harmony} & 19.86/0.7022 & 18.43/0.6099 & 17.52/0.5374 & 25.58/0.7554 & 21.79/0.6675 & 20.61/0.6047 & 16.95/0.4374 & 16.61/0.4854 & 16.57/0.4202 & 19.32/0.5800 \\

    CR-former \cite{wu2024cr} & 18.56/0.7117 & 18.46/0.6792 & 18.30/0.6322 & 22.29/0.7725 & 22.24/0.7699 & 22.13/0.7664 & 17.83/0.7619 & 17.70/0.7277 & 17.46/0.6592 & 19.44/0.7201 \\
    \midrule
    Ours (Direct)  & 18.81/0.6741 & 18.43/0.5878 & 17.68/0.4016 & 22.54/0.7300 & 21.85/0.6558 & 20.09/0.4432 & 17.73/0.6925 & 17.47/0.5903 & 16.77/0.3859 & 19.04/0.5736 \\
    Ours (Sequential)  & \textbf{30.67/0.8489} & \textbf{29.17/0.7968} & \textbf{27.45/0.7158} & \textbf{35.29/0.9000} & \textbf{34.67/0.8901} & \textbf{33.68/0.8766} & \textbf{26.93/0.8198} & \textbf{25.92/0.7784} & \textbf{24.08/0.6982} & \textbf{29.76/0.8138} \\
    \bottomrule
    \end{tabular}%
    }
  \label{tab:compose_degradation}%
\end{table*}%

\begin{table*}[htbp]
  \centering
  \caption{{Quantitative comparison on out-of-distribution three-composite-degradation tasks. Evaluated methods include RSIR baselines, our direct approach (task embedding averaging), and our sequential approach (ordered task-specific subtasks).}}
  \resizebox{.99\textwidth}{!}  {
    \begin{tabular}{m{6.8em}<{\centering}cccccccccc}
    \toprule
    \multirow{2}[2]{*}{Method} & \multicolumn{3}{c}{Denoising+Deblurring+Declouding on RICE1} & \multicolumn{3}{c}{Denoising+Deblurring+Declouding on RICE2} & \multicolumn{3}{c}{Denoising+Deblurring+Deshadowing on SRD} & \multirow{2}[2]{*}{Average} \\
\cmidrule(r){2-4}  \cmidrule(r){5-7} \cmidrule(r){8-10}  
& $\sigma=15$ & $\sigma=25$ & $\sigma=50$ & $\sigma=15$ & $\sigma=25$ & $\sigma=50$ & $\sigma=15$ & $\sigma=25$ & $\sigma=50$ &    \\
    \midrule
    PromptIR \cite{potlapalli2306promptir} & 18.97/0.6514 & 18.61/0.6324 & 18.29/0.6011 & 22.97/0.7831 & 22.50/0.7708 & 22.20/0.7560 & 17.47/0.6483 & 17.32/0.6120 & 17.09/0.5467 & 19.49/0.6690 \\
    AdaIR \cite{cuiadair} & 19.02/0.6622 & 18.64/0.6415 & 18.39/0.6085 & 22.41/0.7773 & 22.30/0.7719 & 22.27/0.7642 & 17.46/0.6594 & 17.32/0.6205 & 17.08/0.5561 & 19.43/0.6735 \\
    MOCE-IR \cite{zamfir2025complexity} & 19.51/0.6572 & 18.96/0.6342 & 18.45/0.5862 & 22.61/0.7744 & 22.24/0.7607 & 22.03/0.7352 & 17.87/0.6543 & 17.44/0.6164 & 17.14/0.5434 & 19.58/0.6624 \\
    Art  \cite{wu2024harmony} & 20.38/0.6778 & 19.90/0.5685 & 18.10/0.4843 & 28.51/0.7493 & 23.45/0.6350 & 20.98/0.5494 & 18.34/0.5145 & 16.99/0.4680 & 16.45/0.3962 & 20.34/0.5603 \\
    CR-former \cite{wu2024cr} & 18.55/0.6720 & 18.44/0.6480 & 18.32/0.6154 & 22.46/0.7756 & 22.38/0.7736 & 22.28/0.7694 & 17.51/0.6649 & 17.31/0.6259 & 17.07/0.5621 & 19.37/0.6785 \\
    \midrule
    Ours (Direct) & 17.84/0.3898 & 17.12/0.2450 & 15.20/0.1020 & 20.51/0.4004 & 19.34/0.2297 & 16.72/0.0876 & 16.90/0.4490 & 16.33/0.3183 & 14.85/0.1685 & 17.20/0.2656 \\
    
    Ours (Sequential) & \textbf{20.98/0.7187} & \textbf{20.39/0.6832} & \textbf{18.90/0.6301} & \textbf{30.55/0.8685} & \textbf{30.24/0.8629} & \textbf{30.03/0.8535} & \textbf{23.01/0.6780} & \textbf{22.54/0.6418} & \textbf{21.62/0.5809} & \textbf{24.25/0.7242} \\
    \bottomrule
    \end{tabular}%
}
  \label{tab:compose_degradation3}%
\end{table*}%

\begin{figure*}[!t]
	\centering
	\includegraphics[width=7in]{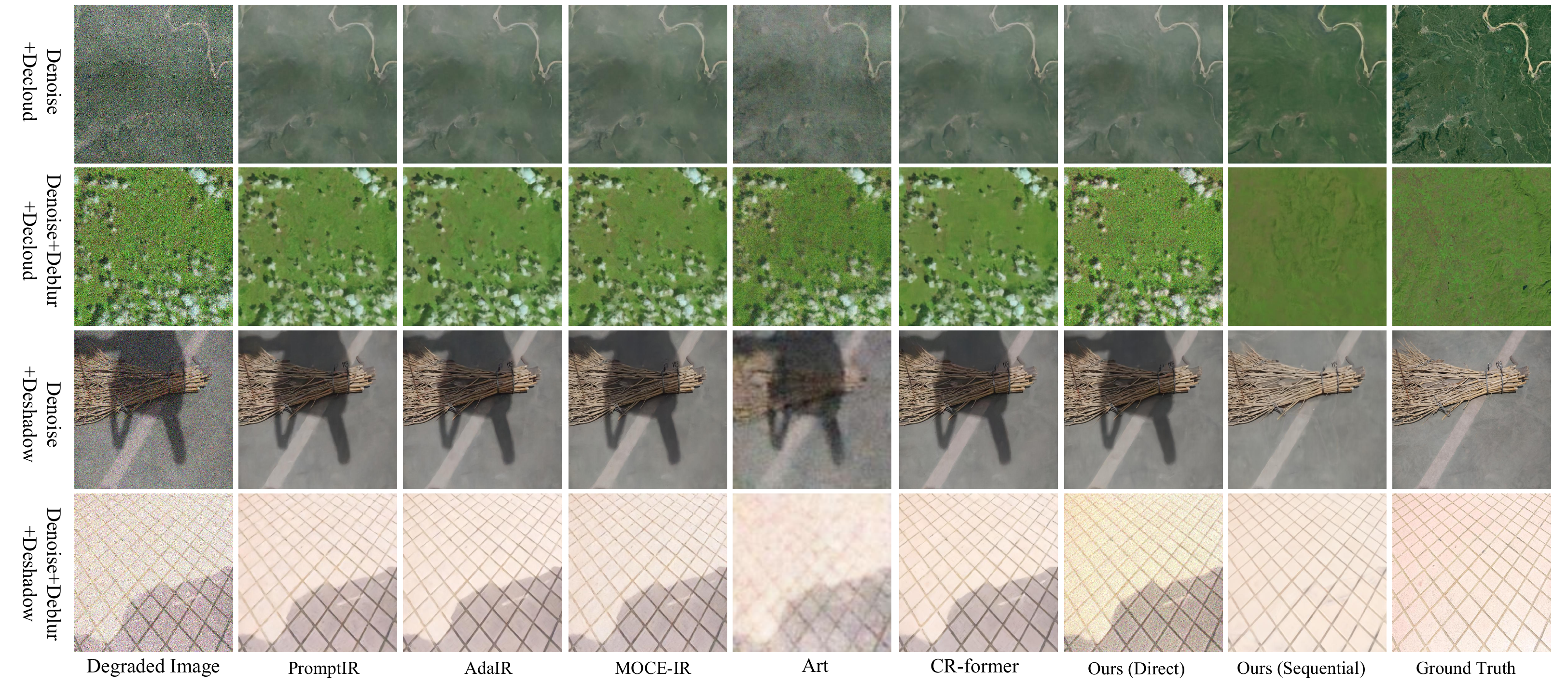}
	\caption{{Visual comparison of restored images on composite degradation tasks under Gaussian noise ( $\sigma=50$)}. Zooming in is recommended for detailed inspection.}
	\label{fig:compose}
\end{figure*}

\subsection{Experimental Analysis of Out-of-Domain Generalization on Composite Degradation Tasks}

This subsection evaluates the out-of-domain generalization capability of our proposed TGPNet and other unified networks on unseen complex composite degradation tasks. All models were trained on our URSIR benchmark and tested on representative composite restoration tasks—Denoising+Declouding, Denoising+Deshadowing, {Denoising+Deblurring+Declouding, and Denoising+Deblurring+Deshadowing}—across three Gaussian noise levels ($\sigma$=15, 25, 50).

\subsubsection{Composite Degradation Generation} 

To construct out-of-domain test cases that mimic real-world remote sensing scenarios (where multiple degradations often coexist but are not included in training data), we followed a two-step process: first, we loaded degradation images from RICE1/RICE2 (for cloud-related composite tasks) and SRD (for shadow-related composite tasks); second, we added Gaussian {blur} or noise to these images to generate composite degraded samples. For example, the input for the Denoising+Declouding task contained both intrinsic cloud cover and artificially added Gaussian noise, reflecting the complexity of real degraded remote sensing imagery.

\subsubsection{Sequential Processing via TGP}

When handling out-of-distribution composite degradations unseen during training, TGPNet exhibits a unique compositional capability. While it cannot directly resolve these unknown combinations effectively, direct methods—such as averaging task embeddings of two individual degradations (e.g., denoising and declouding)—are ineffective (Fig. \ref{fig:compose-step}, column 2; Table \ref{tab:compose_degradation}), as they fail to account for each degradation’s distinct attributes and processing order.

TGPNet addresses this by decomposing composite degradations into sequential task-specific subtasks. For example, in the Denoising+Declouding case (Fig. \ref{fig:compose-step}, Row 1), TGPNet first uses a denoise-specific task vector to remove Gaussian noise; this step significantly suppresses noise while preserving cloud cover. Next, a decloud-specific vector is applied to the denoised intermediate result, eliminating cloud artifacts to produce a high-fidelity image. For Denoising+Deshadowing (Row 2), the same logic applies: noise is removed first, followed by shadow elimination.
This stepwise approach is enabled by TGP’s learnable task embeddings, which dynamically modulate network feature processing to adapt to each degradation type at each step. By prioritizing task-specific modulation in sequence, TGPNet ensures precise restoration even for novel composite degradations, effectively bridging the gap between training-domain and out-of-distribution tasks through structured decomposition.


\subsubsection{Comparison with Baselines} 

Quantitative results in Tables \ref{tab:compose_degradation} {and \ref{tab:compose_degradation3}} confirm TGPNet’s decisive superiority over all baseline methods across all tasks, datasets, and noise levels. For instance, on the RICE1 Denoising+Declouding task with $\sigma$=15, TGPNet sequential processing achieves 30.67 PSNR and 0.8489 SSIM—surpassing the next-best baseline (PromptHSI) by over 8 dB in PSNR. While PromptHSI uses text-guided prompts (e.g., “denoise+decloud") and thus outperforms methods without explicit guidance, it lacks TGP’s hierarchical task decomposition and remains far inferior to TGPNet. In contrast, PromptIR—which initializes prompts randomly at the start of training and lacks explicit task decomposition, exhibits poor out-of-domain performance (similar to non-prompt baselines), confirming that random initialization fails to capture composite degradation patterns. This comparison demonstrates that explicit sequential task guidance (via TGP) is critical for effective composite degradation handling.

\subsubsection{Out-of-Domain Robustness}

TGPNet exhibits unprecedented out-of-distribution generalization capability. While competing baselines struggle with unseen composite tasks, TGPNet maintains outstanding performance. Importantly, TGPNet shows strong robustness to increasing degradation severity: its performance degrades gradually as noise level (\(\sigma\)) increases, whereas baselines experience a sharp performance collapse under high noise. Visual results in Fig. \ref{fig:compose} further validate this advantage: TGPNet produces artifact-free restored images, while baselines leave obvious residual noise, cloud remnants, or uneliminated shadows.

The hierarchical TGP strategy enables dynamic, adaptive feature processing that outperforms fixed-parameterization approaches. By decomposing composite degradations into sequential task-specific steps, TGPNet represents not merely an incremental improvement but a qualitative advance in real-world remote sensing image restoration—where degradations are often composite and unseen during training. This establishes TGPNet as a highly practical solution for scenarios requiring out-of-domain generalization, marking a paradigm shift in multi-degradation image restoration.

\begin{figure*}[!t]
	\centering
	\includegraphics[width=6.5in]{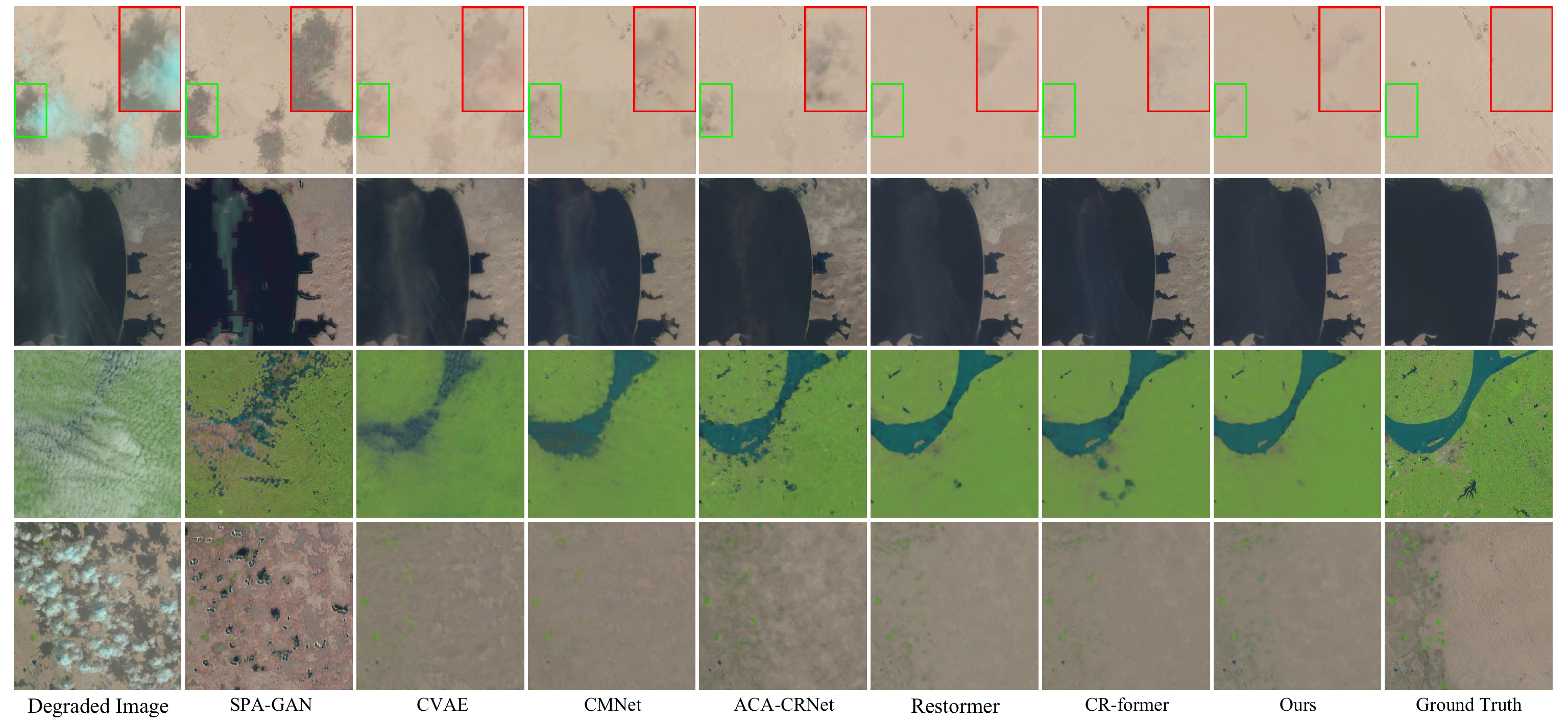}
	\caption{
		Visual comparison of restored images for single-degradation declouding on RICE2. Key local details (in green boxes) are highlighted in red boxes, and zooming in is recommended for detailed inspection.
	}
	\label{fig:performance-1task}
\end{figure*}

\begin{table}[!t]
	\centering
	\caption{Comparisons for single declouding degradation on the RICE2 dataset. Higher PSNR and SSIM indicate better performance, while lower MAE and SAM are preferable. Best in bold.}
	\begin{tabular}{cc
			ccc}
		\toprule
		Method & PSNR$\uparrow$  & SSIM$\uparrow$  & MAE$\downarrow$   & SAM$\downarrow$  \\
		\midrule
		SPA-GAN \cite{pan2020cloud} & 27.51 & 0.8177 & 0.0403 & 3.36  \\
		CVAE \cite{ding2022cvae}  & 33.62 & 0.9079 & 0.0216 & 1.69  \\
		CMNet \cite{liu2024cascaded} & 34.92 & 0.9121 & 0.0188 & 1.40 \\
		ACA-CRNet \cite{huang2024ACAcrnet} & 35.65 & 0.9126 & \textbf{0.0164} & 1.29   \\
		Restormer\cite{zamir2022restormer}  & 35.65 & 0.9124 & 0.0173 & 1.33  \\
		CR-former\cite{wu2024cr}  & 35.61 & 0.9123 & 0.0168 & 1.33 \\
		\midrule
		Ours  & \textbf{35.89} & \textbf{0.9132} & {0.0168} & \textbf{1.27 }  \\
		\bottomrule
	\end{tabular}%
	\label{tab:performance-cloud}%
\end{table}%

\subsection{Performance on Single Declouding Degradation}

We evaluate TGPNet's specialized cloud removal capability on the RICE2 dataset, benchmarking against six state-of-the-art methods: generative models (SPA-GAN \cite{pan2020cloud}, CVAE \cite{ding2022cvae}), attention-based networks (CMNet \cite{liu2024cascaded}, ACA-CRNet \cite{huang2024ACAcrnet}), and transformer architectures (Restormer \cite{zamir2022restormer}, CR-former \cite{wu2024cr}).

\subsubsection{Quantitative Results}

As shown in Table \ref{tab:performance-cloud}, TGPNet achieves state-of-the-art performance. It attains the highest PSNR (35.89 dB) and SSIM (0.9132), alongside the lowest SAM (1.27°), indicating superior signal fidelity and spectral consistency. TGPNet outperforms the attention-based ACA-CRNet by 0.24 dB in PSNR and surpasses transformer-based baselines (Restormer, CR-former) in spectral preservation with a 0.06° improvement in SAM, while maintaining a highly competitive MAE (0.0168).

\subsubsection{Visualization Results}

Fig. \ref{fig:performance-1task} illustrates qualitative comparisons across diverse terrains. TGPNet effectively eliminates thick clouds and associated shadows over soil regions (Rows 1, 4) and removes thin cloud layers over marine scenes without compromising aquatic textures (Row 2). Furthermore, it recovers heavily obscured structural details, such as river courses (Row 3), with minimal artifacts. These results demonstrate TGPNet's robustness in handling variable cloud densities while preserving scene semantics.

Collectively, these findings confirm that TGPNet not only excels in unified multi-task settings but also surpasses specialized models in targeted single-degradation scenarios.

\begin{table}[t]
  \centering
  \caption{Efficiency comparison of TGPNet with unified image restoration models across four key metrics (lower is better). 
  }
\resizebox{.9\columnwidth}{!}  {
    \begin{tabular}{m{7em}<{\centering}m{3.5em}<{\centering}m{3.5em}<{\centering}m{5em}<{\centering}m{3.5em}<{\centering}}
    \toprule
    Method & Params(M) & FLOPs(G) & GPU memory(MB) & Times(ms) \\
    \midrule
    PromptIR \cite{potlapalli2306promptir} & 35.59  & 177.15  & 1753  & 166.21  \\
    AdaIR \cite{cuiadair} & 28.79  & 165.89  & 1735  & 169.31  \\
    MOCE-IR \cite{zamfir2025complexity} & 11.48  & 44.12  & 1007  & 88.65  \\
    PromptHSI \cite{lee2024prompthsi} & 25.91  & 408.58  & 5227  & 218.22  \\
    {Art \cite{wu2024harmony}} & 37.79 &	4.90 & 973 & 7.16 \\ 
    CR-former \cite{wu2024cr}  & 20.86  & 71.22  & 1561  & 72.88  \\
    \midrule
    Ours  & 21.27  & 71.42  & 1643  & 60.84  \\
    \bottomrule
    \end{tabular}%
   }
  \label{tab:efficiency-all}%
\end{table}%

\begin{table}[t]
  \centering
  \caption{ Efficiency comparison of TGPNet with declouding models across four key metrics (lower is better).
  }
\resizebox{.9\columnwidth}{!}  {
    \begin{tabular}{m{7.5em}<{\centering}m{3.5em}<{\centering}m{3.5em}<{\centering}m{5em}<{\centering}m{3.5em}<{\centering}}
    \toprule
    Method & Params(M) & FLOPs(G) & GPU memory(MB) & Times(ms) \\
    \midrule
    SPA-GAN \cite{pan2020cloud} & 0.21  & 15.2  & 3749  & 26.00  \\
    CVAE \cite{ding2022cvae} & 15.42 & 37.1  & 1097  & 15.10  \\
    CMNet \cite{liu2024cascaded} & 16.51 & 236.0  & 1607  & 197.00  \\
    ACA-CRNet \cite{huang2024ACAcrnet} & 20.39 & 1422.0  & 6417  & 223.70  \\
    Restormer \cite{zamir2022restormer} & 26.13 & 155.0  & 1629  & 112.30  \\
    \midrule
    Ours  & 21.27 & 71.4  & 1643  & 60.84  \\
    
    \bottomrule
    \end{tabular}%
}
  \label{tab:efficiency-cloud}%
\end{table}%

\subsection{Efficiency Analysis}

We evaluated the computational efficiency of TGPNet against state-of-the-art models using an NVIDIA RTX 3090 with 
$256 \times 256$ inputs (Tables \ref{tab:efficiency-all} and \ref{tab:efficiency-cloud}).


The results demonstrate that TGPNet occupies a highly favorable position on the efficiency-performance spectrum. With 21.27M parameters and 71.42G FLOPs, our model is substantially more lightweight than large-scale unified models like PromptIR (177.15G FLOPs) and PromptHSI (408.58G FLOPs), as well as computationally intensive specialized methods such as ACA-CRNet (1422G FLOPs). While more compact models like MOCE-IR (11.48M) exist, TGPNet provides a superior performance-to-complexity ratio.

However, TGPNet's most critical advantage is its inference speed. At 60.84 ms, it is the fastest model among all high-performing competitors, outperforming the next fastest model, CR-Former, by a significant 16.5\% and also surpassing specialized models like Restormer. This combination of robust performance, moderate complexity, and state-of-the-art inference speed confirms that TGPNet strikes an effective trade-off, making it highly suitable for large-scale or time-sensitive remote sensing applications.

\begin{table*}[t]
	\centering
	\caption{Ablation study for the proposed Task-Guided Prompt (TGP) module on the URSIR benchmark (PSNR/SSIM; higher = better; best in bold).}
	\resizebox{.99\textwidth}{!}  {
		\begin{tabular}{m{6.5em}<{\centering}cccccccc}
			\toprule
			\multirow{2}[2]{*}{Method} & \multicolumn{2}{c}{Declouding} & \multicolumn{1}{c}{Deshadowing} &  \multicolumn{3}{c}{Denoising on UCMLUD} & \multicolumn{1}{c}{SAR Despeckling} & \multirow{2}[2]{*}{Average} \\
            \cmidrule(r){2-3}  \cmidrule(r){4-4} \cmidrule(r){5-7} \cmidrule(r){8-8} 
            & on RICE1 & on RICE2 & on SRD &  sigma=15 & sigma=25 & sigma=50 &  On NRD &  \\
			\midrule
			baseline  & 35.29/0.9606 & 36.07/0.9136 & 28.19/0.8744 &  34.25/0.9253 & \textbf{31.97}/0.8868 & 29.04/0.8166 & 24.88/0.8284 & 31.38/0.8865 \\
			w/ TGP (Relu) & 35.68/0.9601 & 36.15/0.9143 & \textbf{28.44/0.8758} &  \textbf{34.27/0.9255} & \textbf{31.97/0.8869} & 29.04/0.8162 & \textbf{24.96}/0.8283 & 31.50/0.8867 \\
			w/ TGP (Ours) & \textbf{35.73/0.9608} & \textbf{36.35/0.9145} & {28.40/0.8747} &  \textbf{34.27/0.9255} & \textbf{31.97}/0.8868 & \textbf{29.05/0.8169} & 24.88/\textbf{0.8329} & \textbf{31.52/0.8874}  \\
			\bottomrule
		\end{tabular}%
	}
	\label{tab:ab1}%
\end{table*}%

\begin{figure}[!t]
	\centering
	\includegraphics[width=3.5in]{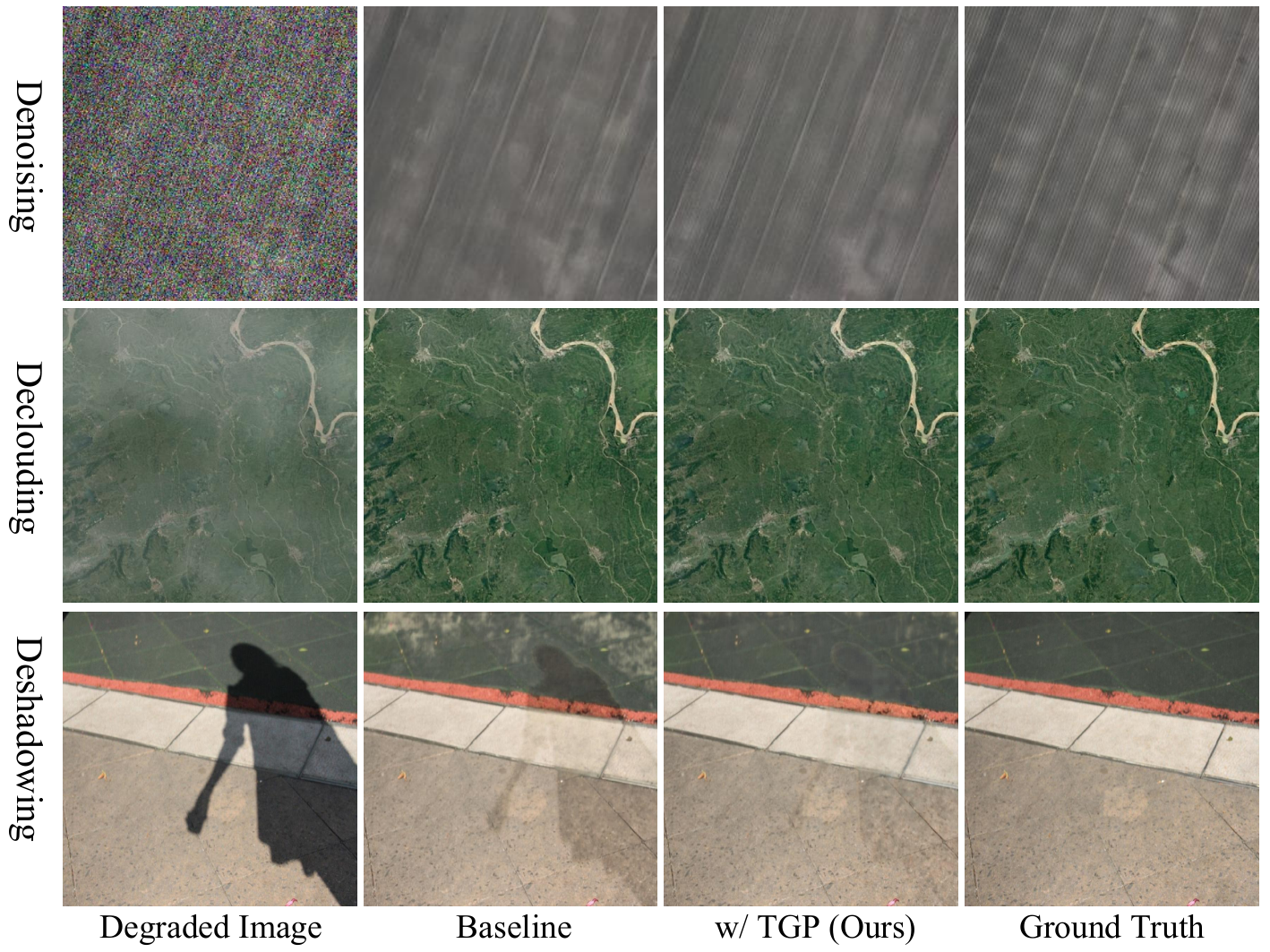}
	\caption{Visualization of ablation study results comparing the baseline model and our TGP-enhanced unified architecture. Zooming in facilitates better visualization and comparison.
	}
	\label{fig:ablation}
\end{figure}

\subsection{Ablation Study}

To assess and validate the impact of the proposed Task-Guided Prompting (TGP) module, we conducted a systematic ablation study using the URSIR benchmark.

\subsubsection{Effectiveness of TGP module}

We evaluated two model configurations: (1) a baseline model, which solely comprises the U-shaped transformer backbone, and (2) the TGPNet model, which incorporates the TGP modules into the baseline. The following analysis provides evidence of the TGP module's efficacy through quantitative, qualitative, and feature-space evaluations.

\paragraph{Quantitative Results}
The quantitative results, presented in Table \ref{tab:ab1}, demonstrate that the TGP module provides consistent and significant performance gains. For declouding, it achieves a 0.44 dB PSNR gain on RICE1 and 0.28 dB on RICE2; for deshadowing, a 0.21 dB PSNR improvement; for denoising, an average 0.01 dB gain across noise levels ($\sigma=15/25/50$); and for SAR despeckling, a 0.0045 SSIM enhancement. Collectively, this amounts to an average PSNR increase of 0.14 dB (from 31.38 to 31.52) and a relative SSIM improvement of 0.009 (from 0.8865 to 0.8874), quantitatively validating that the TGP module is a critical component for enhancing restoration fidelity.


\paragraph{Qualitative Results}

The qualitative results in Fig. \ref{fig:ablation} visually corroborate these quantitative gains. The visual comparison reveals TGPNet's superiority in several key aspects: it preserves finer textures in denoised farmland regions (Row 1), achieves more complete removal of thin clouds without sacrificing ground-truth details (Row 2), and more accurately reconstructs information within shadowed areas (Row 3). This analysis confirms that the TGP module's ability to provide task-specific guidance translates directly to tangible improvements in detail recovery and artifact suppression across a range of remote sensing degradations.


\begin{table}[t]
  \centering
  \caption{Clustering performance of decoder stage 2 features base and with TGP module. A higher metric is better.}
  \resizebox{.99\columnwidth}{!}  {
    \begin{tabular}{m{6.5em}<{\centering}m{3em}<{\centering}m{2.5em}<{\centering}m{2.5em}<{\centering}m{2.5em}<{\centering}m{2.5em}<{\centering}m{2.5em}<{\centering}}
    \toprule
    \multirow{2}[2]{*}{Method} & \multicolumn{3}{c}{Internal Metrics} & \multicolumn{3}{c}{External Metrics} \\
\cmidrule(r){2-4}  \cmidrule(r){5-7} 
& Silhouette & CH    & Dunn    & ARI   & AMI   & FMI \\
    \midrule
    baseline  & 0.1605 & 163.11 & 0.1450 & 0.4837  & 0.6348 & 0.6407 \\
    w/ TGP (Ours) & \textbf{0.4388} & \textbf{572.76} & \textbf{0.2754} & \textbf{0.9610 } & \textbf{0.9279} & \textbf{0.9736} \\
    \bottomrule
    \end{tabular}%
    }
  \label{tab:ab-cluster}%
\end{table}%

\begin{figure}[!t]
	\centering
	\includegraphics[width=3.5in]{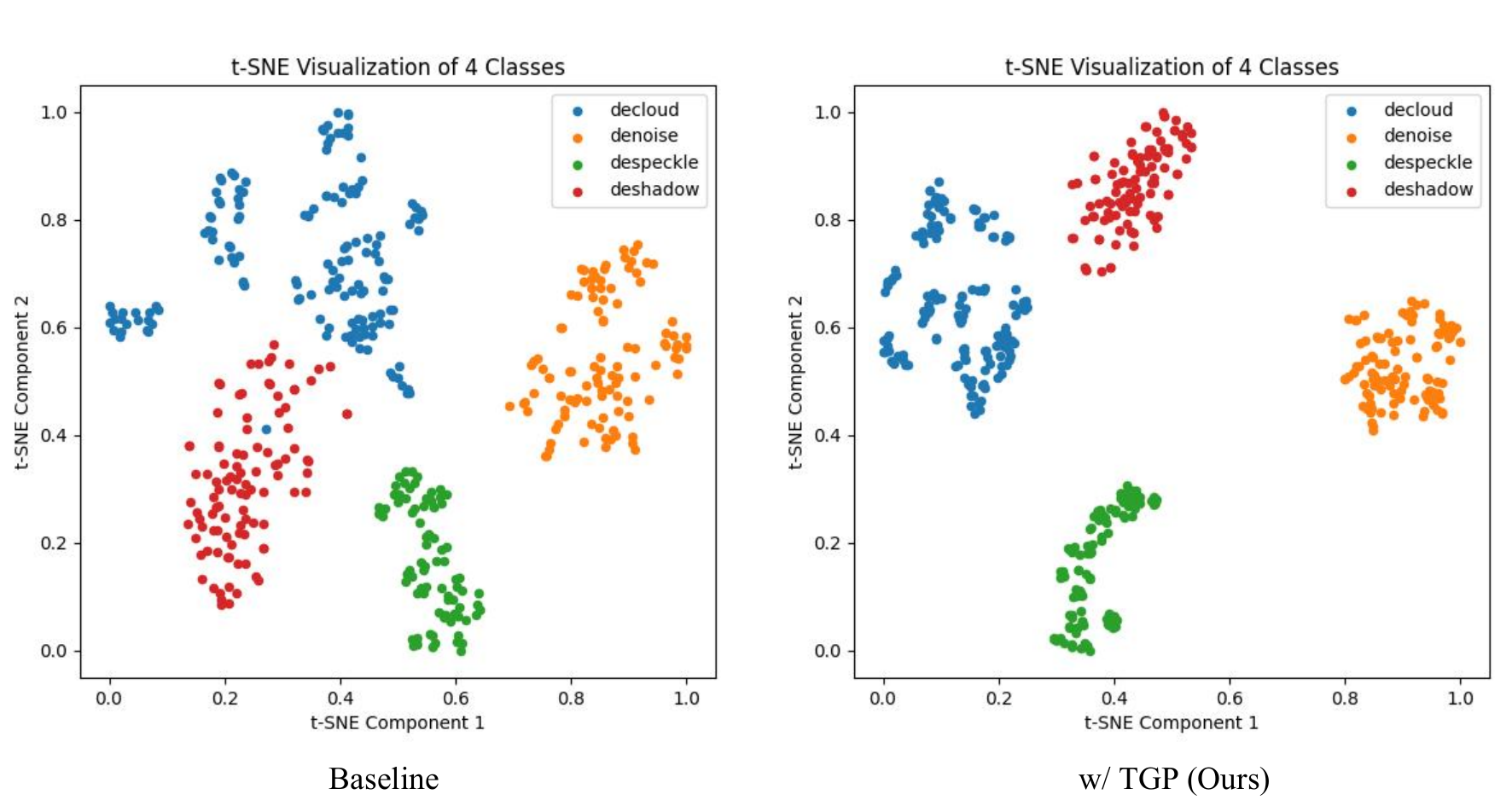}
	\caption{
    t-SNE visualization of decoder stage 2 features for the baseline vs. TGPNet across four degradation tasks.
	}
	\label{fig:abl-tsne}
\end{figure}

\paragraph{Feature Cluster Analysis}

To assess learned representation quality, we performed K-means clustering (K=4) on features from decoder stage 2 ($D_2$) of the baseline and TGPNet. We used internal/external metrics (Table \ref{tab:ab-cluster}) and t-SNE visualization (Fig. \ref{fig:abl-tsne}) for evaluation.

TGPNet’s features are far more discriminative: it outperforms the baseline on all internal metrics (Silhouette Coefficient: 0.4388 vs. 0.1605; CH Index: 572.76 vs. 163.11; Dunn Index: 0.2754 vs. 0.145) and external metrics (ARI: 0.9610 vs. 0.4837; AMI/FMI also show large gains), aligning closely with ground-truth degradation classes. The t-SNE plot confirms this: TGPNet’s features form tight, well-separated clusters for each degradation type, while the baseline’s features are entangled. 

Ablation results show the Task-Guided Prompting (TGP) module is key to TGPNet’s success. It boosts quantitative performance, visual detail, and feature separability—proving the module effectively guides task-specific restoration.

\subsubsection{Justification for Affine Feature Modulation}

We validate the design choice of linear affine modulation through theoretical grounding and empirical ablation.

\paragraph{Theoretical Justification}
The sufficiency of affine modulation is supported by established frameworks like FiLM \cite{perez2018film} and SFT \cite{wang2018recovering}. Affine parameters ($\gamma$, $\beta$) act as feature-wise gates that selectively emphasize task-relevant features by manipulating activation statistics (mean and variance) \cite{wang2018recovering}. While the modulation step is linear, the parameters are generated by the deep, non-linear LTSE module. This ensures the overall system retains high expressivity and can model complex conditional distributions without requiring explicit activation functions within the modulation layer \cite{perez2018film}.

\paragraph{Empirical Verification}
We compared the standard linear modulation against a non-linear variant (adding ReLU post-modulation). As shown in Table \ref{tab:ab1} (Rows 2 vs. 3), the non-linear variant yields negligible performance difference (~0.02 dB). This confirms that the non-linearities inherent in the LTSE and backbone are sufficient, validating the linear formulation as the optimal choice for computational efficiency.

\subsubsection{{Analysis of Learned Task Prompts Before and After TGP Module}}

{To better understand the information captured by each prompt, we visualized and analyzed the learned task prompts before and after the TGP module in the decoder across four and five degradation tasks.}

\begin{table}[!t]
	\centering
	\caption{{Clustering performance of decoder stage 2 features before and after TGP module in both 4-task and 5-task settings. A higher metric is better.}}
	\resizebox{.99\columnwidth}{!}  {
	\begin{tabular}{m{2.0em}<{\centering}m{5em}<{\centering}m{3em}<{\centering}m{2.5em}<{\centering}m{2.5em}<{\centering}m{2.5em}<{\centering}m{2.5em}<{\centering}m{2.5em}<{\centering}}
		\toprule
		\multirow{2}[2]{*}{Tasks} & \multirow{2}[2]{*}{Method} & \multicolumn{3}{c}{Internal Metrics} & \multicolumn{3}{c}{External Metrics} \\
		\cmidrule(r){3-5}  \cmidrule(r){6-8}          &       & Silhouette & CH    & Dunn  & ARI   & AMI   & FMI \\
		\midrule
		\multirow{2}[2]{*}{4-task} & before TGP & 0.1621 & 209.77 & 0.1649 & 0.4669 & 0.6359 & 0.626 \\
		& {after TGP} & \textbf{0.1623} & \textbf{237.75} & \textbf{0.2046} & \textbf{0.5206} & \textbf{0.6714} & \textbf{0.6634} \\
		\cmidrule{2-8}    \multirow{2}[2]{*}{5-task} & before TGP & 0.3448 & \textbf{649.51} & 0.1908 & 0.707 & 0.784 & 0.7783 \\
		& {after TGP} & \textbf{0.4314} & 522.41 & \textbf{0.312} & \textbf{0.9124} & \textbf{0.8842} & \textbf{0.9338} \\
		\bottomrule
	\end{tabular}%
}
	\label{tab:ab_before_after_tgp}%
\end{table}%

\begin{figure}[!t]
	\centering
	\includegraphics[width=3.5in]{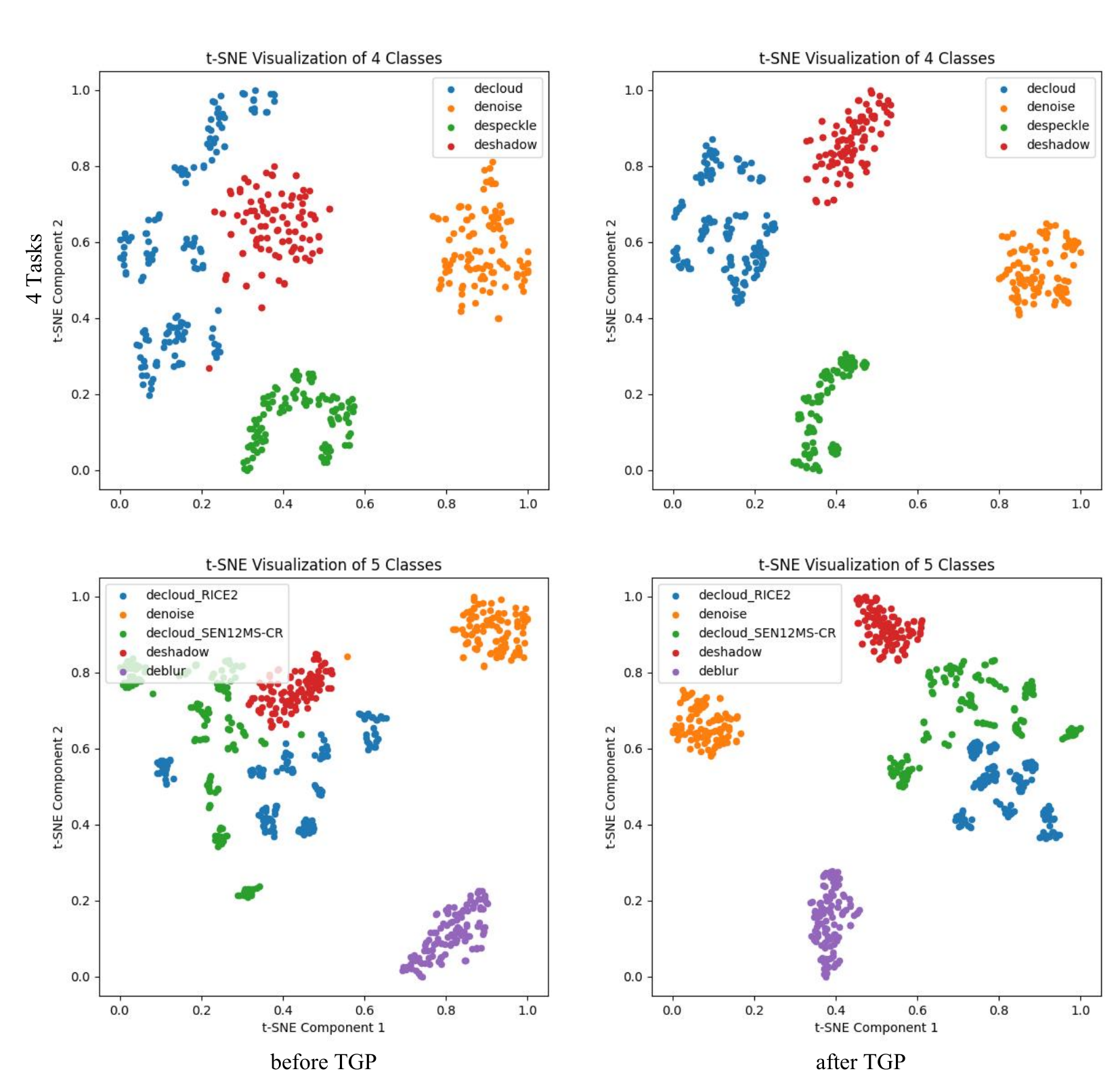}
	\caption{
		{t-SNE visualization of decoder stage 2 features before and after application of the TGP module, shown for both 4-task and 5-task settings.}
	}
	\label{fig:ab_before_after_tgp_tsne}
\end{figure}

\paragraph{Feature Disentanglement and Visualization}
{The t-SNE visualizations in Fig. \ref{fig:ab_before_after_tgp_tsne} clearly demonstrate the TGP module's ability to disentangle task-related features. Before the TGP module, the features are highly entangled; for instance, representations for declouding and deshadowing are largely indistinguishable. After the TGP module processes these features, they form distinct, well-separated clusters corresponding to each restoration task. Notably, even similar tasks like declouding on RGB and multispectral images are resolved into separate, yet proximate, clusters, reflecting their semantic relationship.}

\paragraph{Quantitative Cluster Analysis}
{This visual evidence is supported by a quantitative analysis using K-means clustering, with the results summarized in Table \ref{tab:ab_before_after_tgp}. The TGP module yields a significant improvement across nearly all internal and external clustering metrics. The most compelling results are seen in the external metrics, which measure alignment with ground-truth task labels. For the five-task scenario, the Adjusted Rand Index (ARI) jumps from 0.707 to 0.9124, and the Fowlkes-Mallows Index (FMI) increases from 0.7783 to 0.9338. These near-perfect scores confirm that the TGP module effectively transforms ambiguous features into discriminative representations that are accurately aligned with their corresponding tasks.}

{Collectively, this analysis demonstrates that the TGP module is highly effective at refining representations to make them task-specific. This ability to learn disentangled features is crucial for enabling a single, unified model to handle diverse restoration tasks.}

\subsubsection{{Comparison with Language-based Prompts}}

{We compared our randomly initialized learnable prompts against prompts initialized with language features from a pre-trained CLIP model. The results in Table \ref{tab:5task-clip} reveal a clear modality-dependent trade-off: language-based prompts offer a slight advantage on most optical (RGB-centric) tasks, but our randomly initialized prompts perform better on multispectral declouding (SEN12MS-CR: 29.61 vs. 29.50 dB PSNR) and SAR despeckling (NRD: 24.95 vs. 24.93 dB PSNR). This suggests that while semantic priors benefit RGB tasks, adapting prompts to RS-specific modalities (multispectral, SAR) is critical for comprehensive restoration. A hybrid method combining both strategies remains a promising direction for future work.}

\begin{table*}[htbp]
	\centering
	\caption{{Comparison of prompt strategies across five RS degradation types (PSNR/SSIM; higher = better; best in bold).}}
	\resizebox{.99\textwidth}{!}  {
		\begin{tabular}{ccccccccc}
			\toprule
			\multirow{2}[2]{*}{Method} & \multicolumn{3}{c}{Declouding} & \multicolumn{1}{l}{Deshadowing} & Denoising & \multicolumn{1}{l}{SAR Despeckling} &  Deblurring & \multirow{2}[2]{*}{Average} \\
			\cmidrule(r){2-4}  \cmidrule(r){5-5} \cmidrule(r){6-6} \cmidrule(r){7-7} \cmidrule(r){8-8}         & On RICE1 & On RICE2 & On SEN12MS-CR & On SRD  & Avg. on UCMLUD & On NRD & Avg. on HIT-UAV &  \\
			
			\midrule
			Random learnable prompts (Ours) & 35.59/0.9591 & {35.93/{0.9130}} & \textbf{29.61}/0.8963 & 27.97/\textbf{0.8739} & {31.55/0.8723} & \textbf{24.95}/0.8294 & {32.35/0.8789} & {31.14/0.8890} \\
			\midrule
			Language-based prompts (CLIP) & \textbf{35.76/0.9601} & \textbf{36.22/0.9131} & 29.50/\textbf{0.8973} & \textbf{28.31}/0.8736 & \textbf{31.60/0.8733} & 24.93/\textbf{0.8312} &  \textbf{32.41/0.8797} & \textbf{31.25/0.8898} \\
			\bottomrule
		\end{tabular}%
	}
	\label{tab:5task-clip}%
\end{table*}%


\subsubsection{{Analysis of SAR Despeckling Performance Limitation}}

{TGPNet’s SAR despeckling PSNR trade-off stems from its additive residual design ($I_{out} = I_{in} + I_r$), which is ill-suited for SAR’s speckle noise. This structural bias leads residual feature learning to prioritize additive tasks (e.g., declouding, Gaussian denoising) that dominate training data, limiting SAR-specific feature capacity.
	Table \ref{tab:train_iterations} validates this: while average task PSNR rises consistently, SAR despeckling PSNR peaks at 300k iterations then declines as residual space is reallocated to additive tasks. Critically, SAR SSIM improves throughout training—confirming the residual network’s strength in preserving structural integrity, which offsets pixel-level limitations for practical SAR use.
	Future work will enhance residual features via task-specific adapters (for multiplicative noise patterns) and orthogonality constraints (to reduce interference), fully resolving the PSNR gap while retaining TGPNet’s multi-task efficiency.}

\begin{table}[!t]
	\centering
	\caption{{Performance comparison between Baseline and TGPNet at different training iterations (PSNR/SSIM; higher = better; best in bold).}}
	\begin{tabular}{m{3em}<{\centering}m{5em}<{\centering}m{5em}<{\centering}m{5em}<{\centering}m{5em}<{\centering}}
		\toprule
		\multirow{2}[2]{*}{\makecell{Train \\  Iterations}} & \multicolumn{2}{c}{Baseline} & \multicolumn{2}{c}{TGPNet} \\
		\cmidrule(lr){2-3} \cmidrule(lr){4-5}       & Avg. on 4 Tasks & SAR Despeckling & Avg. on 4 Tasks & SAR Despeckling \\
		\midrule
		200,000 & 30.72/0.8844 & \textbf{25.08}/0.8275 & 31.32/0.8856 & 24.90/0.8278 \\
		\rowcolor[gray]{0.9}
		300,000 & 31.27/0.8855 & 25.03/0.8309 & 31.35/0.8859 & \textbf{24.96}/0.8291 \\
		400,000 & 31.28/0.8859 & 24.94/0.8267 & 31.51/0.8874 & 24.85/0.8324 \\
		500,000 & \textbf{31.38/0.8865} & 24.88/\textbf{0.8284} & \textbf{31.52/0.8874} & 24.88/\textbf{0.8329} \\
		\bottomrule
	\end{tabular}%
	\label{tab:train_iterations}%
\end{table}%

\section{Conclusion}

This work presents TGPNet, a unified framework designed to tackle the challenge of multi-degradation restoration in remote sensing imagery. By leveraging a Task-Guided Prompting (TGP) mechanism, the framework enables a single network to dynamically adapt its features for diverse tasks across distinct sensor modalities (optical, SAR, and TIR). Extensive benchmarking demonstrates that TGPNet achieves performance competitive with specialized state-of-the-art models, validating the feasibility of parameter-efficient, unified restoration architectures.

However, two primary limitations warrant discussion. First, the framework currently relies on explicit task priors (non-blind restoration), constraining its deployment in real-world scenarios where degradation types are unknown. Second, joint optimization across fundamentally distinct imaging domains (e.g., coherent SAR speckle vs. additive optical noise) remains challenging; without robust regularization, the shared backbone faces potential negative transfer due to conflicting optimization dynamics.
Future research will aim to overcome these constraints by integrating self-supervised degradation estimation for blind unified restoration and exploring lightweight, modality-specific adapters to mitigate domain conflicts.

\bibliographystyle{IEEEtran}
\bibliography{IEEEabrv,IEEEexample}

\begin{thebibliography}{10}
\providecommand{\url}[1]{#1}
\csname url@samestyle\endcsname
\providecommand{\newblock}{\relax}
\providecommand{\bibinfo}[2]{#2}
\providecommand{\BIBentrySTDinterwordspacing}{\spaceskip=0pt\relax}
\providecommand{\BIBentryALTinterwordstretchfactor}{4}
\providecommand{\BIBentryALTinterwordspacing}{\spaceskip=\fontdimen2\font plus
\BIBentryALTinterwordstretchfactor\fontdimen3\font minus
  \fontdimen4\font\relax}
\providecommand{\BIBforeignlanguage}[2]{{%
\expandafter\ifx\csname l@#1\endcsname\relax
\typeout{** WARNING: IEEEtran.bst: No hyphenation pattern has been}%
\typeout{** loaded for the language `#1'. Using the pattern for}%
\typeout{** the default language instead.}%
\else
\language=\csname l@#1\endcsname
\fi
#2}}
\providecommand{\BIBdecl}{\relax}
\BIBdecl

\bibitem{casagli2023landslide}
N.~Casagli, E.~Intrieri, V.~Tofani, G.~Gigli, and F.~Raspini, ``Landslide
  detection, monitoring and prediction with remote-sensing techniques,''
  \emph{Nature Reviews Earth \& Environment}, vol.~4, no.~1, pp. 51--64, 2023.

\bibitem{sheffield2018satellite}
J.~Sheffield, E.~F. Wood, M.~Pan, H.~Beck, G.~Coccia, A.~Serrat-Capdevila, and
  K.~Verbist, ``Satellite remote sensing for water resources management:
  Potential for supporting sustainable development in data-poor regions,''
  \emph{Water Resources Research}, vol.~54, no.~12, pp. 9724--9758, 2018.

\bibitem{holloway2018statistical}
J.~Holloway and K.~Mengersen, ``Statistical machine learning methods and remote
  sensing for sustainable development goals: A review,'' \emph{Remote Sensing},
  vol.~10, no.~9, p. 1365, 2018.

\bibitem{zakria2022multiscale}
Z.~Zakria, J.~Deng, R.~Kumar, M.~S. Khokhar, J.~Cai, and J.~Kumar, ``Multiscale
  and direction target detecting in remote sensing images via modified
  yolo-v4,'' \emph{IEEE Journal of Selected Topics in Applied Earth
  Observations and Remote Sensing}, vol.~15, pp. 1039--1048, 2022.

\bibitem{kotaridis2021remote}
I.~Kotaridis and M.~Lazaridou, ``Remote sensing image segmentation advances: A
  meta-analysis,'' \emph{ISPRS Journal of Photogrammetry and Remote Sensing},
  vol. 173, pp. 309--322, 2021.

\bibitem{wang2024rsid}
Z.~Wang, X.~He, B.~Xiao, L.~Chen, and X.~Bi, ``Rsid-cr: Remote sensing image
  denoising based on contrastive learning,'' \emph{IEEE Journal of Selected
  Topics in Applied Earth Observations and Remote Sensing}, 2024.

\bibitem{wu2024cr}
Y.~Wu, Y.~Deng, S.~Zhou, Y.~Liu, W.~Huang, and J.~Wang, ``Cr-former: Single
  image cloud removal with focused taylor attention,'' \emph{IEEE Transactions
  on Geoscience and Remote Sensing}, 2024.

\bibitem{liu2024cascaded}
J.~Liu, B.~Pan, and Z.~Shi, ``Cascaded memory network for optical remote
  sensing imagery cloud removal,'' \emph{IEEE Transactions on Geoscience and
  Remote Sensing}, vol.~62, pp. 1--11, 2024.

\bibitem{guo2023shadowformer}
L.~Guo, S.~Huang, D.~Liu, H.~Cheng, and B.~Wen, ``Shadowformer: Global context
  helps shadow removal,'' in \emph{Proceedings of the AAAI conference on
  artificial intelligence}, vol.~37, no.~1, 2023, pp. 710--718.

\bibitem{xiao2024homoformer}
J.~Xiao, X.~Fu, Y.~Zhu, D.~Li, J.~Huang, K.~Zhu, and Z.-J. Zha, ``Homoformer:
  Homogenized transformer for image shadow removal,'' in \emph{Proceedings of
  the IEEE/CVF conference on computer vision and pattern recognition}, 2024,
  pp. 25\,617--25\,626.

\bibitem{ko2021sar}
J.~Ko and S.~Lee, ``Sar image despeckling using continuous attention module,''
  \emph{IEEE Journal of Selected Topics in Applied Earth Observations and
  Remote Sensing}, vol.~15, pp. 3--19, 2021.

\bibitem{fang2024contrastive}
Y.~Fang, R.~Liu, Y.~Peng, J.~Guan, D.~Li, and X.~Tian, ``Contrastive learning
  for real sar image despeckling,'' \emph{ISPRS Journal of Photogrammetry and
  Remote Sensing}, vol. 218, pp. 376--391, 2024.

\bibitem{li2022all}
B.~Li, X.~Liu, P.~Hu, Z.~Wu, J.~Lv, and X.~Peng, ``All-in-one image restoration
  for unknown corruption,'' in \emph{Proceedings of the IEEE/CVF conference on
  computer vision and pattern recognition}, 2022, pp. 17\,452--17\,462.

\bibitem{potlapalli2306promptir}
V.~Potlapalli, S.~Zamir, S.~Khan, and F.~Khan, ``Promptir: Prompting for
  all-in-one blind image restoration,'' \emph{arXiv preprint arXiv:2306.13090},
  vol.~6.

\bibitem{cuiadair}
Y.~Cui, S.~W. Zamir, S.~Khan, A.~Knoll, M.~Shah, and F.~S. Khan, ``Adair:
  Adaptive all-in-one image restoration via frequency mining and modulation,''
  in \emph{The Thirteenth International Conference on Learning
  Representations}.

\bibitem{rasti2021image}
B.~Rasti, Y.~Chang, E.~Dalsasso, L.~Denis, and P.~Ghamisi, ``Image restoration
  for remote sensing: Overview and toolbox,'' \emph{IEEE Geoscience and Remote
  Sensing Magazine}, vol.~10, no.~2, pp. 201--230, 2021.

\bibitem{shen2022coupling}
H.~Shen, M.~Jiang, J.~Li, C.~Zhou, Q.~Yuan, and L.~Zhang, ``Coupling model-and
  data-driven methods for remote sensing image restoration and fusion:
  Improving physical interpretability,'' \emph{IEEE Geoscience and Remote
  Sensing Magazine}, vol.~10, no.~2, pp. 231--249, 2022.

\bibitem{xu2018deep}
W.~Xu, G.~Xu, Y.~Wang, X.~Sun, D.~Lin, and Y.~Wu, ``Deep memory connected
  neural network for optical remote sensing image restoration,'' \emph{Remote
  Sensing}, vol.~10, no.~12, p. 1893, 2018.

\bibitem{hu2024hybrid}
S.~Hu, F.~Gao, X.~Zhou, J.~Dong, and Q.~Du, ``Hybrid convolutional and
  attention network for hyperspectral image denoising,'' \emph{IEEE Geoscience
  and Remote Sensing Letters}, vol.~21, pp. 1--5, 2024.

\bibitem{jin2025mb}
Z.~Jin, Y.~Qiu, K.~Zhang, H.~Li, and W.~Luo, ``Mb-taylorformer v2: improved
  multi-branch linear transformer expanded by taylor formula for image
  restoration,'' \emph{IEEE Transactions on Pattern Analysis and Machine
  Intelligence}, 2025.

\bibitem{zhang2021deep}
K.~Zhang, R.~Li, Y.~Yu, W.~Luo, and C.~Li, ``Deep dense multi-scale network for
  snow removal using semantic and depth priors,'' \emph{IEEE Transactions on
  Image Processing}, vol.~30, pp. 7419--7431, 2021.

\bibitem{zhang2022enhanced}
K.~Zhang, D.~Li, W.~Luo, W.~Ren, and W.~Liu, ``Enhanced spatio-temporal
  interaction learning for video deraining: faster and better,'' \emph{IEEE
  Transactions on Pattern Analysis and Machine Intelligence}, vol.~45, no.~1,
  pp. 1287--1293, 2022.

\bibitem{zhang2018adversarial}
K.~Zhang, W.~Luo, Y.~Zhong, L.~Ma, W.~Liu, and H.~Li, ``Adversarial
  spatio-temporal learning for video deblurring,'' \emph{IEEE Transactions on
  Image Processing}, vol.~28, no.~1, pp. 291--301, 2018.

\bibitem{wang2025lldiffusion}
T.~Wang, K.~Zhang, Y.~Zhang, W.~Luo, B.~Stenger, T.~Lu, T.-K. Kim, and W.~Liu,
  ``Lldiffusion: Learning degradation representations in diffusion models for
  low-light image enhancement,'' \emph{Pattern Recognition}, vol. 166, p.
  111628, 2025.

\bibitem{mullissa2020despecknet}
A.~G. Mullissa, D.~Marcos, D.~Tuia, M.~Herold, and J.~Reiche, ``Despecknet:
  Generalizing deep learning-based sar image despeckling,'' \emph{IEEE
  Transactions on Geoscience and Remote Sensing}, vol.~60, pp. 1--15, 2020.

\bibitem{pang2024hir}
L.~Pang, X.~Rui, L.~Cui, H.~Wang, D.~Meng, and X.~Cao, ``Hir-diff: Unsupervised
  hyperspectral image restoration via improved diffusion models,'' in
  \emph{Proceedings of the IEEE/CVF Conference on Computer Vision and Pattern
  Recognition}, 2024, pp. 3005--3014.

\bibitem{feng2024progressive}
Y.~Feng, Y.~Yang, X.~Fan, Z.~Zhang, L.~Bu, and J.~Zhang, ``A progressive image
  restoration network for high-order degradation imaging in remote sensing,''
  \emph{arXiv preprint arXiv:2412.07195}, 2024.

\bibitem{lee2024prompthsi}
C.-M. Lee, C.-H. Cheng, Y.-F. Lin, Y.-C. Cheng, W.-T. Liao, C.-C. Hsu, F.-E.
  Yang, and Y.-C.~F. Wang, ``Prompthsi: Universal hyperspectral image
  restoration framework for composite degradation,'' \emph{arXiv e-prints}, pp.
  arXiv--2411, 2024.

\bibitem{jiang2024survey}
J.~Jiang, Z.~Zuo, G.~Wu, K.~Jiang, and X.~Liu, ``A survey on all-in-one image
  restoration: Taxonomy, evaluation and future trends,'' \emph{arXiv preprint
  arXiv:2410.15067}, 2024.

\bibitem{chen2021pre}
H.~Chen, Y.~Wang, T.~Guo, C.~Xu, Y.~Deng, Z.~Liu, S.~Ma, C.~Xu, C.~Xu, and
  W.~Gao, ``Pre-trained image processing transformer,'' in \emph{Proceedings of
  the IEEE/CVF conference on computer vision and pattern recognition}, 2021,
  pp. 12\,299--12\,310.

\bibitem{ai2024lora}
Y.~Ai, H.~Huang, and R.~He, ``Lora-ir: taming low-rank experts for efficient
  all-in-one image restoration,'' \emph{arXiv preprint arXiv:2410.15385}, 2024.

\bibitem{zamfir2025complexity}
E.~Zamfir, Z.~Wu, N.~Mehta, Y.~Tan, D.~P. Paudel, Y.~Zhang, and R.~Timofte,
  ``Complexity experts are task-discriminative learners for any image
  restoration,'' in \emph{Proceedings of the Computer Vision and Pattern
  Recognition Conference}, 2025, pp. 12\,753--12\,763.

\bibitem{guo2024onerestore}
Y.~Guo, Y.~Gao, Y.~Lu, H.~Zhu, R.~W. Liu, and S.~He, ``Onerestore: A universal
  restoration framework for composite degradation,'' in \emph{European
  conference on computer vision}.\hskip 1em plus 0.5em minus 0.4em\relax
  Springer, 2024, pp. 255--272.

\bibitem{mao2024allrestorer}
J.~Mao, Y.~Yang, X.~Yin, L.~Shao, and H.~Tang, ``Allrestorer: All-in-one
  transformer for image restoration under composite degradations,'' \emph{arXiv
  preprint arXiv:2411.10708}, 2024.

\bibitem{ozdenizci2023restoring}
O.~Ozdenizci and R.~Legenstein, ``Restoring vision in adverse weather
  conditions with patch-based denoising diffusion models,'' \emph{IEEE
  Transactions on Pattern Analysis \& Machine Intelligence}, vol.~45, no.~08,
  pp. 10\,346--10\,357, 2023.

\bibitem{ai2024multimodal}
Y.~Ai, H.~Huang, X.~Zhou, J.~Wang, and R.~He, ``Multimodal prompt perceiver:
  Empower adaptiveness generalizability and fidelity for all-in-one image
  restoration,'' in \emph{Proceedings of the IEEE/CVF Conference on Computer
  Vision and Pattern Recognition}, 2024, pp. 25\,432--25\,444.

\bibitem{jiang2024autodir}
Y.~Jiang, Z.~Zhang, T.~Xue, and J.~Gu, ``Autodir: Automatic all-in-one image
  restoration with latent diffusion,'' in \emph{European Conference on Computer
  Vision}.\hskip 1em plus 0.5em minus 0.4em\relax Springer, 2024, pp. 340--359.

\bibitem{chen2025unirestore}
I.~Chen, W.-T. Chen, Y.-W. Liu, Y.-C. Chiang, S.-Y. Kuo, M.-H. Yang
  \emph{et~al.}, ``Unirestore: Unified perceptual and task-oriented image
  restoration model using diffusion prior,'' in \emph{Proceedings of the
  Computer Vision and Pattern Recognition Conference}, 2025, pp.
  17\,969--17\,979.

\bibitem{mandal2025unicorn}
D.~Mandal, S.~Chattopadhyay, G.~Tong, and P.~Chakravarthula, ``Unicorn: Latent
  diffusion-based unified controllable image restoration network across
  multiple degradations,'' \emph{arXiv preprint arXiv:2503.15868}, 2025.

\bibitem{zhou2025q}
Y.~Zhou, J.~Cao, Z.~Zhang, F.~Wen, Y.~Jiang, J.~Jia, X.~Liu, X.~Min, and
  G.~Zhai, ``Q-agent: Quality-driven chain-of-thought image restoration agent
  through robust multimodal large language model,'' \emph{arXiv preprint
  arXiv:2504.07148}, 2025.

\bibitem{zeng2025vision}
H.~Zeng, X.~Wang, Y.~Chen, J.~Su, and J.~Liu, ``Vision-language gradient
  descent-driven all-in-one deep unfolding networks,'' in \emph{Proceedings of
  the Computer Vision and Pattern Recognition Conference}, 2025, pp.
  7524--7533.

\bibitem{conde2024instructir}
M.~V. Conde, G.~Geigle, and R.~Timofte, ``Instructir: High-quality image
  restoration following human instructions,'' in \emph{European Conference on
  Computer Vision}.\hskip 1em plus 0.5em minus 0.4em\relax Springer, 2024, pp.
  1--21.

\bibitem{qi2024spire}
C.~Qi, Z.~Tu, K.~Ye, M.~Delbracio, P.~Milanfar, Q.~Chen, and H.~Talebi,
  ``Spire: Semantic prompt-driven image restoration,'' in \emph{European
  Conference on Computer Vision}.\hskip 1em plus 0.5em minus 0.4em\relax
  Springer, 2024, pp. 446--464.

\bibitem{wen2025multi}
Y.~Wen, T.~Gao, J.~Zhang, Z.~Li, and T.~Chen, ``Multi-axis prompt and
  multi-dimension fusion network for all-in-one weather-degraded image
  restoration,'' in \emph{Proceedings of the AAAI Conference on Artificial
  Intelligence}, vol.~39, no.~8, 2025, pp. 8323--8331.

\bibitem{zamir2022restormer}
S.~W. Zamir, A.~Arora, S.~Khan, M.~Hayat, F.~S. Khan, and M.-H. Yang,
  ``Restormer: Efficient transformer for high-resolution image restoration,''
  in \emph{Proceedings of the IEEE/CVF conference on computer vision and
  pattern recognition}, 2022, pp. 5728--5739.

\bibitem{perez2018film}
E.~Perez, F.~Strub, H.~De~Vries, V.~Dumoulin, and A.~Courville, ``Film: Visual
  reasoning with a general conditioning layer,'' in \emph{Proceedings of the
  AAAI conference on artificial intelligence}, vol.~32, no.~1, 2018.

\bibitem{zhao2016loss}
H.~Zhao, O.~Gallo, I.~Frosio, and J.~Kautz, ``Loss functions for image
  restoration with neural networks,'' \emph{IEEE Transactions on computational
  imaging}, vol.~3, no.~1, pp. 47--57, 2016.

\bibitem{yang2010bag}
Y.~Yang and S.~Newsam, ``Bag-of-visual-words and spatial extensions for
  land-use classification,'' in \emph{Proceedings of the 18th SIGSPATIAL
  international conference on advances in geographic information systems},
  2010, pp. 270--279.

\bibitem{lin2019remote}
D.~Lin, G.~Xu, X.~Wang, Y.~Wang, X.~Sun, and K.~Fu, ``A remote sensing image
  dataset for cloud removal,'' \emph{arXiv preprint arXiv:1901.00600}, 2019.

\bibitem{meraner2020cloud}
A.~Meraner, P.~Ebel, X.~X. Zhu, and M.~Schmitt, ``Cloud removal in sentinel-2
  imagery using a deep residual neural network and sar-optical data fusion,''
  \emph{ISPRS Journal of Photogrammetry and Remote Sensing}, vol. 166, pp.
  333--346, 2020.

\bibitem{qu2017deshadownet}
L.~Qu, J.~Tian, S.~He, Y.~Tang, and R.~W. Lau, ``Deshadownet: A multi-context
  embedding deep network for shadow removal,'' in \emph{Proceedings of the IEEE
  conference on computer vision and pattern recognition}, 2017, pp. 4067--4075.

\bibitem{guan2023robust}
J.~Guan, R.~Liu, X.~Tian, X.~Tang, and S.~Li, ``Robust sar image despeckling by
  deep learning from near-real datasets,'' \emph{IEEE Journal of Selected
  Topics in Applied Earth Observations and Remote Sensing}, vol.~17, pp.
  2963--2979, 2023.

\bibitem{suo2023hit}
J.~Suo, T.~Wang, X.~Zhang, H.~Chen, W.~Zhou, and W.~Shi, ``Hit-uav: A
  high-altitude infrared thermal dataset for unmanned aerial vehicle-based
  object detection,'' \emph{Scientific Data}, vol.~10, no.~1, p. 227, 2023.

\bibitem{loshchilov2017decoupled}
I.~Loshchilov and F.~Hutter, ``Decoupled weight decay regularization,''
  \emph{arXiv preprint arXiv:1711.05101}, 2017.

\bibitem{huang2024ACAcrnet}
W.~Huang, Y.~Deng, Y.~Wu, and J.~Wang, ``Attentive contextual attention for
  cloud removal,'' \emph{IEEE Transactions on Geoscience and Remote Sensing},
  2024.

\bibitem{wu2024harmony}
G.~Wu, J.~Jiang, K.~Jiang, and X.~Liu, ``Harmony in diversity: Improving
  all-in-one image restoration via multi-task collaboration,'' in
  \emph{Proceedings of the 32nd ACM international conference on multimedia},
  2024, pp. 6015--6023.

\bibitem{luo2020deeply}
S.~Luo, H.~Li, and H.~Shen, ``Deeply supervised convolutional neural network
  for shadow detection based on a novel aerial shadow imagery dataset,''
  \emph{ISPRS Journal of Photogrammetry and remote sensing}, vol. 167, pp.
  443--457, 2020.

\bibitem{pan2020cloud}
H.~Pan, ``Cloud removal for remote sensing imagery via spatial attention
  generative adversarial network,'' \emph{arXiv preprint arXiv:2009.13015},
  2020.

\bibitem{ding2022cvae}
H.~Ding, Y.~Zi, and F.~Xie, ``Uncertainty-based thin cloud removal network via
  conditional variational autoencoders,'' in \emph{Proceedings of the Asian
  Conference on Computer Vision}, 2022, pp. 469--485.

\bibitem{wang2018recovering}
X.~Wang, K.~Yu, C.~Dong, and C.~C. Loy, ``Recovering realistic texture in image
  super-resolution by deep spatial feature transform,'' in \emph{Proceedings of
  the IEEE conference on computer vision and pattern recognition}, 2018, pp.
  606--615.

\end{thebibliography}

\vspace{-33pt}
\begin{IEEEbiography}[{\includegraphics[width=1in,height=1.25in,clip,keepaspectratio]{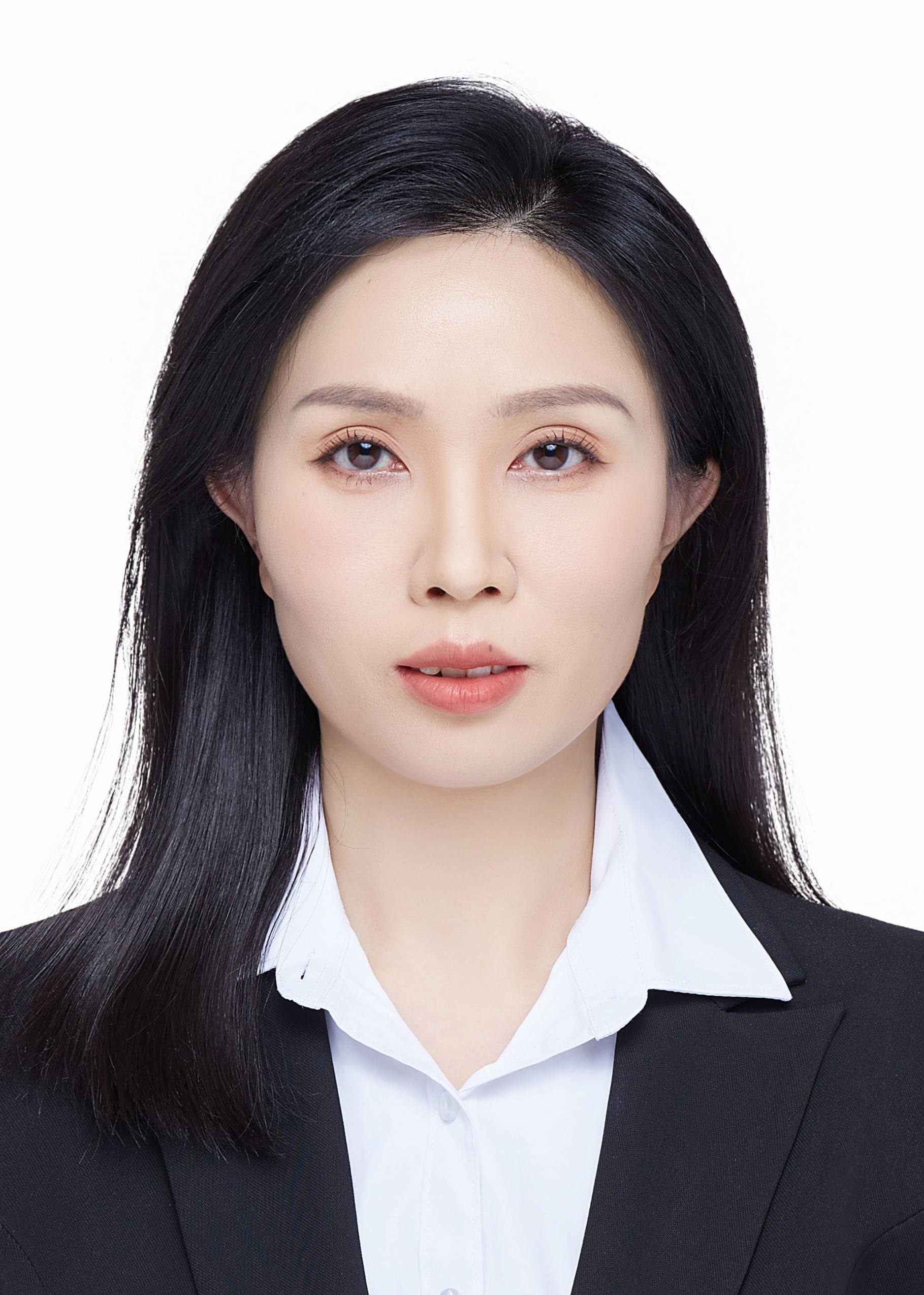}}]{Wenli Huang} received her Ph.D. from Xi'an Jiaotong University, Xi'an, China, in 2023. She is currently a lecturer at Ningbo University of Technology. Her research interests include deep learning, image processing, and computer vision, focusing on network structure optimization, image inpainting, image restoration, etc.
\end{IEEEbiography}

\vspace{-33pt}
\begin{IEEEbiography}[{\includegraphics[width=1in,height=1.25in,clip,keepaspectratio]{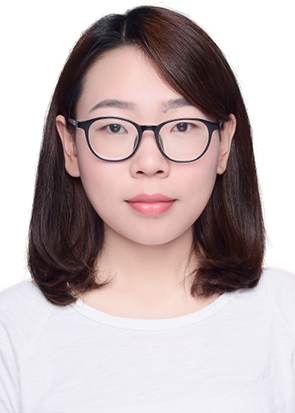}}]{Yang Wu} received the M.E. degree from Northwestern Polytechnical University, Xi’an, China, in 2018. She is currently pursuing a Ph.D. degree in the Institute of Artificial Intelligence and Robotics at Xi’an Jiaotong University. Her research interests include Knowledge graph completion and Graph Representation Learning.
\end{IEEEbiography}

\vspace{-33pt}
\begin{IEEEbiography}[{\includegraphics[width=1in,height=1.25in,clip,keepaspectratio]{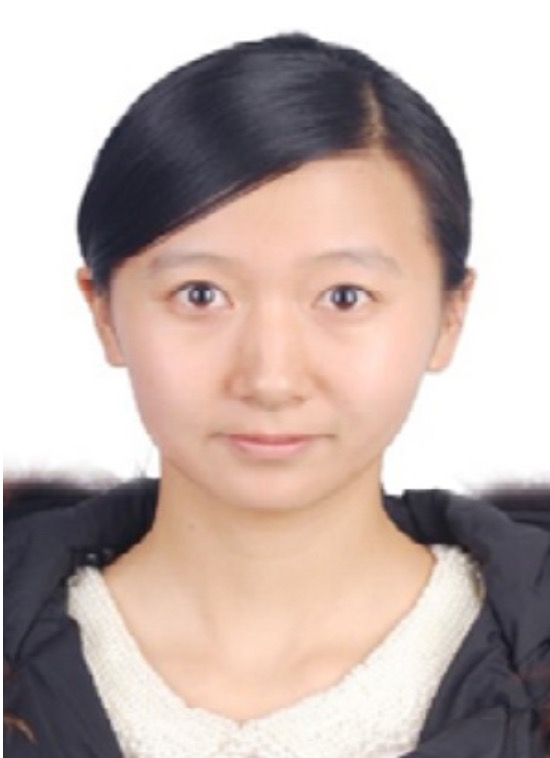}}]{Xiaomeng Xin} received the M.E. degree from ShaanXi Normal University, Xi an, China, in 2015. She is currently pursuing the Ph.D. degree in Institute of Artificial Intelligence and Robotics at Xian Jiaotong University. Her research interests include person re-identification, incremental learning and image restoration, etc.
\end{IEEEbiography}

\vspace{-33pt}
\begin{IEEEbiography}[{\includegraphics[width=1.1in,height=1.25in,clip,keepaspectratio]{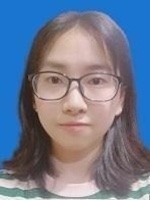}}]{Zhihong Liu} received the M.E. degree from Xi'an Jiaotong University, Xi'an, China, in 2023. She is currently pursuing a Ph.D. degree in the Institute of Artificial Intelligence and Robotics at Xi'an Jiaotong University. Her research interests include scene understanding and multimodal representation learning.
\end{IEEEbiography}

\vspace{-33pt}
\begin{IEEEbiography}[{\includegraphics[width=1in,height=1.25in,clip,keepaspectratio]{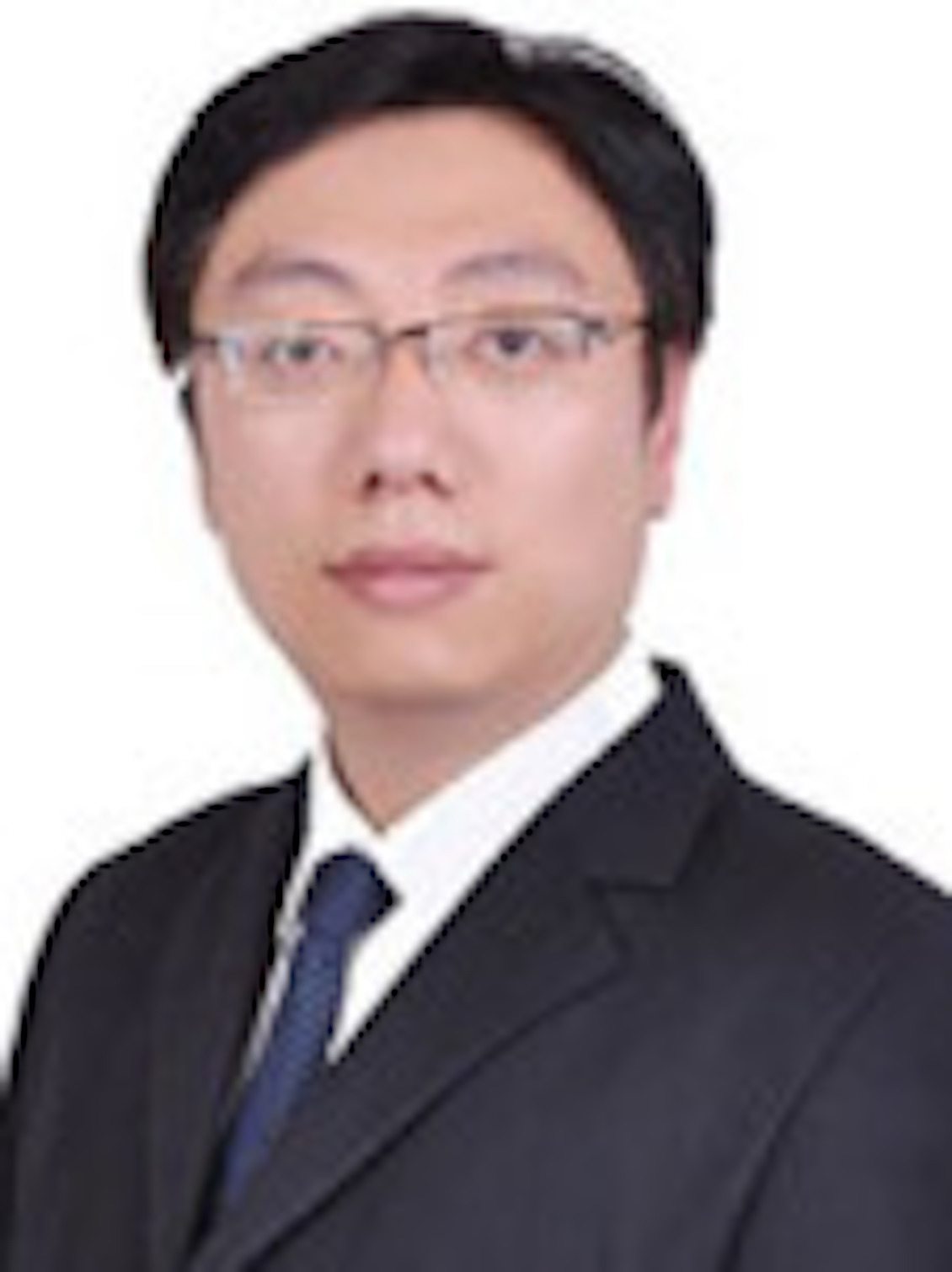}}]{Jinjun Wang} received the B.E. and M.E. degrees from the Huazhong University of Science and Technology, China, in 2000 and 2003, respectively. He received his Ph.D. from Nanyang Technological University, Singapore, in 2006. From 2006 to 2009, he was with NEC Laboratories America, Inc., as a Research Scientist. From 2010 to 2013, he was with Epson Research and Development, Inc., as a Senior Research Scientist. He is currently a Professor at Xi'an Jiaotong University. His research interests include pattern classification, image/video enhancement and editing, content-based image/video annotation and retrieval, semantic event detection, etc.
\end{IEEEbiography}

\vspace{-33pt}
\begin{IEEEbiography}[{\includegraphics[width=1.1in,height=1.25in,clip,keepaspectratio]{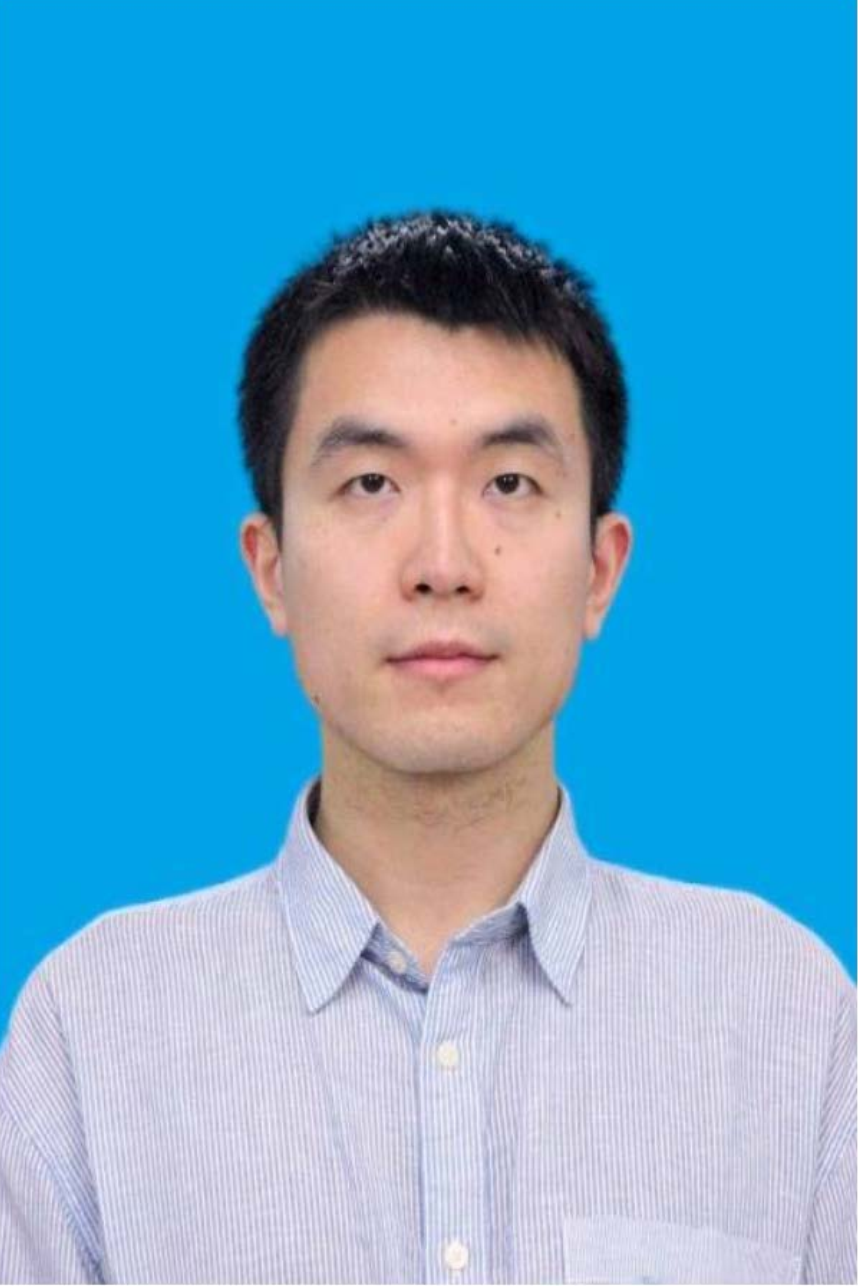}}]{Ye Deng} received his Ph.D. from Xi'an Jiaotong University, Xi'an, China, in 2023. He is currently an Assistant Professor at Southwestern University of Finance and Economics. His research interests include image inpainting, image restoration, and machine learning.
\end{IEEEbiography}




\vspace{11pt}


\vfill

\end{document}